\documentclass[pra,twocolumn,10pt,aps,floatfix]{revtex4-1}
\usepackage[dvips]{graphicx}
\usepackage{epsfig}

\usepackage[margin=2cm,head=0.5cm]{geometry}
\usepackage[latin1]{inputenc}
\usepackage{bm}
\usepackage{amsmath}
\usepackage{amssymb}
\usepackage{latexsym}
\usepackage{amsfonts}
\usepackage{epsfig}
\usepackage{color}
\usepackage[linktocpage, colorlinks=true ,linkcolor=blue, citecolor=blue]{hyperref}
\usepackage[all]{hypcap}

\usepackage{orcidlink}

\newcommand{\expval}[1]{\left< #1 \right>}

\newcommand{\ket}[1]{\left|#1\right>}
\newcommand{\bra}[1]{\left<#1\right|}

\newcommand{\nn}{\nonumber\\}

\newcommand{\f}[1]{\mbox{\boldmath$#1$}}

\newcommand{\bea}{\begin{eqnarray}}
\newcommand{\eea}{\end{eqnarray}}
\newcommand{\beann}{\begin{eqnarray*}}
\newcommand{\eeann}{\end{eqnarray*}}

\newcommand{\trace}[1]{{\rm Tr}\left\{ #1 \right\}}
\newcommand{\traceB}[1]{{\rm Tr_B}\left\{ #1 \right\}}

\newcommand{\abs}[1]{{\left| #1 \right|}}

\newcommand{\ii}{\mathrm{i}}  

\newcommand{\new}[1]{{#1}}
\definecolor{blau}{RGB}{0,70,180}

\begin{document}
  
\title{Performance boost of a collective qutrit refrigerator}

\author{Dmytro Kolisnyk\,\orcidlink{0000-0002-8612-8202}$^{1,2}$}
\author{Gernot Schaller\,\orcidlink{0000-0003-0062-9944}$^{2}$}
\email{g.schaller@hzdr.de}
\affiliation{$^1$Jacobs University Bremen, Campus Ring 1, 28759 Bremen, Germany}
\affiliation{$^2$Helmholtz-Zentrum Dresden-Rossendorf, Bautzner Landstra\ss e 400, 01328 Dresden, Germany}

\date{\today}

\begin{abstract}
A single qutrit with transitions selectively driven by weakly-coupled reservoirs can implement one of the world's smallest refrigerators.
We analyze the performance of $N$ such fridges that are collectively coupled to the reservoirs.
We observe a quantum boost, manifest in a quadratic scaling of the steady-state cooling current with $N$.
As $N$ grows further, the scaling reduces to linear, since the transitions responsible for the quantum boost become energetically unfavorable.
Fine-tuned inter-qutrit interactions may be used to maintain the quantum boost for all $N$ and also for not-perfectly collective scenarios.
\end{abstract}

\maketitle


Beyond foundational questions, the study of quantum systems is nowadays driven by applications of quantum computation~\cite{nielsen2000}.
While for these background noise~\cite{breuer2002} is usually detrimental, it can also be turned into something useful by considering energy conversion processes~\cite{alicki1979a}.
The ability of quantum systems to act as energy filters between reservoirs has fostered the whole field of quantum thermodynamics leading to numerous applications~\cite{binder2019}.
One of the most pressing questions in this area is whether quantum heat engines can outperform their classical counterparts in some aspects~\cite{scully2003b,jaramillo2016a,latune2019a,myers2022a}.

Already on a much simpler setting, the collective superradiant evolution of quantum many-body systems~\cite{dicke1954a,gross1982a,benedict1996a} is  an example where quantum systems can easily surpass the speed of classical ones.
\new{This can for example be exploited in the collective charging of quantum batteries~\cite{ferraro2018a,mayo2022a,ueki2022a}.
Also} finite-stroke quantum thermodynamic cycles have been analyzed with collective quantum working fluids~\cite{hardal2015a,uzdin2016a,cakmak2016a,niedenzu2018a,gelbwaser_klimovsky2019a,kloc2019a,watanabe2020a,kamimura2022a,da_silva_souza2022a}.
These mimic classical thermodynamic cycles like e.g. the Quantum Otto cycle~\cite{kieu2004a,kosloff2017a} and thereby depend on classical control parameters.
Experimentally, such cycle implementations require a high degree of control that should not open additional decoherence channels
and at present dwarfs the work that can be extracted (see e.g.~\cite{rossnagel2016a}). 
The study of autonomously operated thermodynamic cycles~\cite{marchegiani2016a,seah2018a,alicki2021a,strasberg2021a} is so far limited to rather low-dimensional systems, as it requires understanding of non-linear dynamics.

In contrast, continuously operating heat engines~\cite{kosloff2014a} operate while simultaneously coupled to two or more reservoirs at all times.
Examples for such devices are thermoelectric generators~\cite{brantut2013a,whitney2014a,sothmann2014a,um2022a} or quantum absorption refrigerators (QARs)~\cite{linden2010a,levy2012b}.
With regard to collective effects, the latter are particularly interesting~\cite{manzano2019a}, as QARs require only reservoirs without particle exchange, for which superradiance is well-established.
Proposals~\cite{mitchison2016a,erdman2018a,mitchison2019a} and actual realizations~\cite{maslennikov2019a} for QARs exist, and also collective couplings have been implemented experimentally in the past~\cite{leroux2010a,schleier_smith2010a,dalla_torre2013a}.
Therefore, in a previous work~\cite{kloc2021a}, a superradiant QAR constructed from interacting qubits has been analyzed.
While superradiant cooling performance could be observed at steady state, the device required fine-tuned interactions between the qubits throughout and the preparation of an entangled initial state.

In this paper, we consider a simpler realization based on qutrits that -- at least in the simplest variant -- need not interact directly and do not require an entangled initial state.
We start by introducing the model in Sec.~\ref{SEC:model}, then explain the methods in Sec.~\ref{SEC:methods} and discuss our results in Sec.~\ref{SEC:results} before concluding in Sec.~\ref{SEC:conclusions}.
For the interested reader, technical background information is provided in the appendices thereafter.
Throughout the paper, we use units with $\hbar=1$ and $k_{\rm B}=1$, plot only dimensionless quantities, 
use an overbar to denote steady-state quantities, bold symbols to refer to the interaction picture, and $\{\hat A,\hat B\}=\hat A \hat B + \hat B \hat A$ to denote the anti-commutator.


\section{Model}\label{SEC:model}

\begin{figure}[t]
\begin{tabular}{lr}
\includegraphics[width=0.15\textwidth,clip=true]{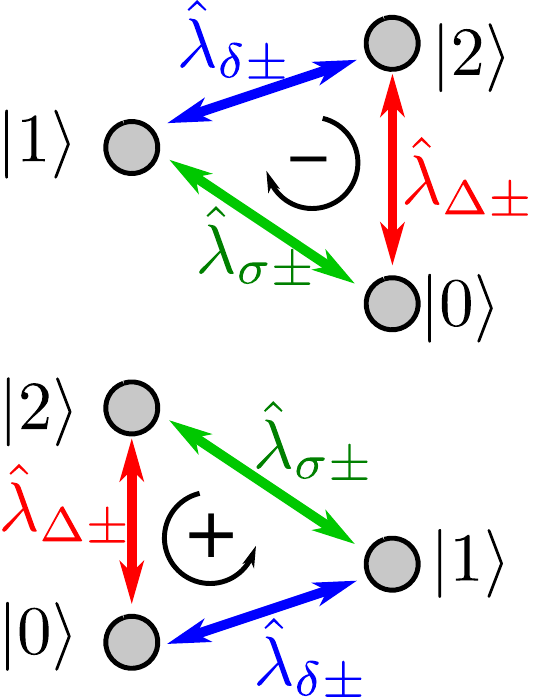} & \includegraphics[width=0.3\textwidth,clip=true]{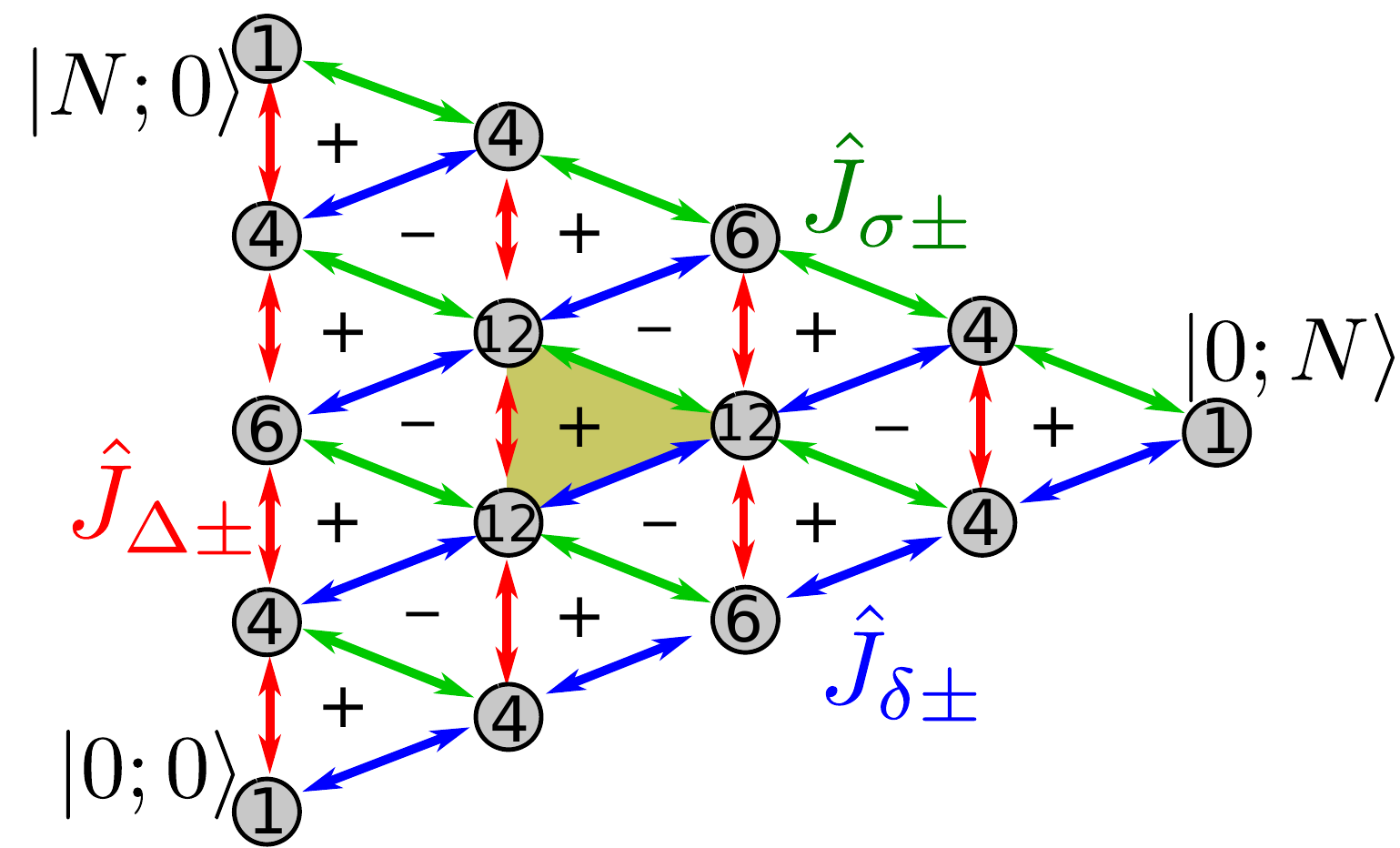}
\end{tabular}
\caption{\label{FIG:threels}
Left: Three-level cyclic systems performing refrigeration ($N=1$, even cycle in bottom left panel and odd cycle with work and cold reservoir exchanged in top left panel).
Right: For larger qutrit numbers (here $N=4$), the $3^N$ states in total are composed of $N!/[(N-M-m)! M! m!]$ states (numbers inside circles) with $M$ large (along red arrows) and $m$ small (along blue arrows) excitations such that $0 \le m+M \le N$.
Ordering the eigenstates in layers by the eigenvalue of Casimir operators, the largest layer corresponds to the subset of permutationally completely symmetric states, which hosts $(N+1)(N+2)/2$ states.
Subspaces of permutationally non-symmetric states can be grouped in additional layers, see Fig.~\ref{FIG:threels4tot}, such that collective couplings then admit only intra-layer transitions.
}
\end{figure}

Our generic starting point is a total Hamiltonian of the form 
\begin{align}\label{EQ:hamgeneral}
    \hat H = \hat{H}_S + \sum_\nu \hat{S}^\nu \otimes \hat{B}^\nu + \sum_\nu \hat{H}_B^\nu
\end{align}
that is composed of a system part $\hat{H}_S$, different reservoirs $\hat{H}_B^\nu$ and the corresponding interactions written as products of bath operators $\hat{B}^\nu$ that couple to different system operators $\hat{S}^\nu$.
In this paper, we will consider three reservoirs $\nu\in\{c,h,w\}$ that we label cold, hot, and work (hottest) reservoirs, respectively.
These reservoirs are modeled by standard harmonic oscillator baths $\hat{H}_B^\nu = \sum_k \omega_{k\nu} \hat{b}_{k\nu}^\dagger \hat{b}_{k\nu}$ throughout.

An introductory example for the system could be a single three-level model (qutrit) with ground state $\ket{0}$, first excited state $\ket{1}$, and most excited state $\ket{2}$, for which we could write 
$\hat{H}_S = \Delta \ket{2}\bra{2} + \delta \ket{1}\bra{1}$, 
where $\Delta > \delta > 0$ are the excitation energies and we have gauged the ground state energy to zero.
We consider the case where the coupling between system and reservoir
drives the individual system transitions exclusively, e.g. for the single qutrit via 
$\hat{S}^c = [\ket{0}\bra{1}+\ket{1}\bra{0}]$,
$\hat{S}^h = [\ket{0}\bra{2}+\ket{2}\bra{0}]$, and
$\hat{S}^w = [\ket{1}\bra{2}+\ket{2}\bra{1}]$, compare Fig.~\ref{FIG:threels} bottom left panel.
In the appropriate regime (reservoirs held in thermal states with inverse temperatures obeying $\beta_c > \beta_h > \beta_w$ and the cooling condition $\beta_h \Delta > \beta_w (\Delta-\delta) + \beta_c \delta$), the single-qutrit model implements a quantum absorption refrigerator (QAR): 
A device that implements stochastic cooling by on average absorbing heat from the coldest ($c$) and hottest ($w$) reservoirs and dumping the waste heat into the intermediate temperature ($h$) reservoir.
We provide more details on the working principles of this configuration and also for the case where the work and cold reservoirs are exchanged (as in Fig.~\ref{FIG:threels} top left panel)
in App.~\ref{APP:qar1}.

In this paper, we will instead as system consider $N$ identical qutrits 
\begin{align}\label{EQ:hamsys}
\hat{H}_S &= \Delta \hat{N}_\Delta + \delta \hat{N}_\delta\,,
\end{align}
where $\hat{N}_\Delta = \sum_{i=1}^N \left(\ket{2}\bra{2}\right)_i$ and $\hat{N}_\delta = \sum_{i=1}^N \left(\ket{1}\bra{1}\right)_i$ count the total number of large and small excitations present in the system, and the introductory example is reproduced for $N=1$.
The reservoirs are assumed to drive individual transitions as before, but the coupling to the $i$th qutrit may in principle depend on its position.
Thus, we assume as system coupling operators the multi-qutrit operators
\begin{align}\label{EQ:coupling}
\hat{S}^c &= \sum_i \left[h_i^c \left(\ket{1}\bra{0}\right)_i + {\rm h.c.}\right] \equiv \hat{S}^c_+ + \hat{S}^c_-\,,\nn
\hat{S}^h &= \sum_i \left[h_i^h\left(\ket{2}\bra{0}\right)_i + {\rm h.c.}\right] \equiv
\hat{S}^h_+ + \hat{S}^h_-\,,\nn
\hat{S}^w &= \sum_i \left[h_i^w \left(\ket{2}\bra{1}\right)_i + {\rm h.c.}\right] \equiv
\hat{S}^w_+ + \hat{S}^w_-\,,
\end{align}
where $\hat{S}^\nu_- = (\hat{S}^\nu_+)^\dagger$.
The dependence of the coupling on the qutrit is encoded in the coefficients $h_i^\nu\in\mathbb{C}$, and furtheron we will denote the limit \new{$h_i^\nu\to 1$} as the {\em collective limit}, for which we write $\hat{J}^\nu = \lim\limits_{h_i^\nu\to 1} \hat{S}^\nu$ and
$\hat{J}^\nu_\pm = \lim\limits_{h_i^\nu\to 1} \hat{S}^\nu_\pm$. 
Analogous to permutationally invariant many-qubit systems, where one can generalize single-qubit Pauli matrices -- the generators of $su(2)$ -- to large spin operators (see e.g.~\cite{wang2002a} for potential applications), we can also define the collective generalizations of the generators of $su(3)$.
We denote them by lower indices 
\begin{align}\label{EQ:collective}
    \hat{J}_\alpha = \frac{1}{2}\sum_{i=1}^N \hat{\lambda}_i^\alpha\,,
\end{align}
where $\hat{\lambda}_i^\alpha$ denotes the Gell-Mann matrix $\hat{\lambda}^\alpha$ (with $1 \le \alpha \le 8$) acting on the $i$th qutrit (with $1\le i \le N$).
In the collective limit, we can express the system coupling operators by the large qutrit operators, specifically we have 
$\hat{S}_\pm^c \to \hat{J}_{\pm}^c = \hat{J}_6\pm\ii \hat{J}_7$, 
$\hat{S}_\pm^h \to \hat{J}_{\pm}^h = \hat{J}_4 \pm \ii \hat{J}_5$, and 
$\hat{S}_\pm^w \to \hat{J}_{\pm}^w = \hat{J}_1\pm\ii \hat{J}_2$,
see App.~\ref{APP:gellmann}.
To represent the problem efficiently, it is advantageous to use common eigenstates of $\hat{J}_3$, $\hat{J}_8$ and the collective Casimir operators of $su(3)$ like $\hat{C}_2 = \sum_{\alpha=1}^8 \hat{J}_\alpha^2$.
We provide some example states in App.~\ref{APP:examples}.

The question we address here is whether one can observe superradiant performance boosts~\cite{yadin2022a} analogous to results for interacting qubits~\cite{manzano2019a,kloc2021a} also for this system of non-interacting qutrits.


\section{Methods}\label{SEC:methods}

We aim at the perturbative treatment of the system-reservoir interaction (i.e., the $\hat{B}^\nu$ operators) and a description of the system by master equations.
Depending on the microscopic implementation of~\eqref{EQ:hamgeneral}, a non-perturbative treatment may require modifications to the global Hamiltonianian~\cite{nataf2010a}.
As in~\eqref{EQ:hamsys} all qutrits are identical and do not interact, the transformation into the interaction picture is straightforward as shown in Eq.~\eqref{EQ:intpic}.
This facilitates the derivation of these master equations.

First, we consider the Redfield-II master equation~\cite{redfield1965} (Lamb-shift omitted, see App.~\ref{APP:redfield} for details)
\begin{align}\label{EQ:redfieldmain}
\dot\rho &= -\ii \left[\hat{H}_S, \rho(t)\right]\nn
&\qquad+\sum_\nu\frac{\gamma_\nu(-\Omega_\nu)}{2} 
\left(\left[\hat{S}^\nu_+ \rho, \hat{S}^\nu\right]+\left[\hat{S}^\nu, \rho \hat{S}^\nu_-\right]\right)\nn
&\qquad+\sum_\nu\frac{\gamma_\nu(+\Omega_\nu)}{2} 
\left(\left[\hat{S}^\nu, \rho \hat{S}^\nu_+\right]+\left[\hat{S}^\nu_-\rho, \hat{S}^\nu\right]\right)\,,
\end{align}
where $\gamma_\nu(\omega)=\Gamma_\nu(\omega)[1+n_\nu(\omega)]\ge 0$ is the product of the reservoir spectral coupling density $\Gamma_\nu(\omega)$ and the Bose distribution $n_\nu(\omega)=[e^{\beta_\nu \omega}-1]^{-1}$, evaluated at the system excitation frequencies $\Omega_c=\delta>0$, $\Omega_h=\Delta>0$, and $\Omega_w=\Delta-\delta>0$.
Since we analytically continue the spectral coupling density as an odd function $\Gamma_\nu(-\omega)=-\Gamma_\nu(+\omega)$, we can express $\gamma_\nu(-\Omega_\nu)=\Gamma_\nu(\Omega_\nu) n_\nu(\Omega_\nu)\ge 0$.
The generator~\eqref{EQ:redfieldmain} need in general not preserve the positivity of the density matrix~\cite{breuer2002} and need not be thermodynamically consistent~\cite{thingna2012a}.
\new{For example, it may not exactly conserve the sum of all stationary energy currents leaving the reservoirs for non-vanishing couplings, such artifacts are of higher order than the accuracy Redfield approach though, cf. Fig.~\ref{FIG:firstlawviolation} in App.~\ref{APP:redfield}.}
Nevertheless, it has for selective systems been shown to approach the true quantum dynamics in the appropriate regimes very well~\cite{hartmann2020a,landi2022a}, such that we use it as benchmark approach here.

Second, we consider the Lindblad-Gorini-Kossakowski-Sudarshan (LGKS) master equation~\cite{lindblad1976a,gorini1976a} (Lamb-shift neglected, see App.~\ref{APP:lindblad} for details)
\begin{align}\label{EQ:lindblad}
\dot{\rho} &= -\ii \left[\hat{H}_S, \rho(t)\right]\nn
&\qquad+\sum_\nu \gamma_\nu(+\Omega_\nu)\left[\hat{S}^\nu_- \rho \hat{S}^\nu_+ - \frac{1}{2}\left\{\hat{S}^\nu_+ \hat{S}^\nu_-, \rho\right\}\right]\nn
&\qquad+\sum_\nu \gamma_\nu(-\Omega_\nu)\left[\hat{S}^\nu_+ \rho \hat{S}^\nu_- - \frac{1}{2}\left\{\hat{S}^\nu_- \hat{S}^\nu_+, \rho\right\}\right]\,,
\end{align}
which unconditionally preserves all density matrix properties and is thermodynamically consistent also for non-equilibrium reservoirs~\cite{schaller2014}.

\new{Third, in the perfectly collective limit ($\hat S^\nu_\pm \to \hat J^\nu_\pm$) and for a completely symmetric initial state like e.g. 
\begin{align}\label{EQ:psirepvac}
    {\ket{\Psi_{\rm rep}^{\rm vac}} =\ket{0}\otimes \ldots \otimes \ket{0} \equiv \ket{0;0}}
\end{align} 
the perfect permutational symmetry is preserved (i.e., formally the evolution is constrained to the subspace with largest Casimir operator eigenvalue). 
The other permutationally completely symmetric states of this subspace with $M$ large and $m$ small excitations can be obtained by acting with  
the collective raising operators on the vacuum state
$\ket{M;m} \propto (J^h_+)^M (J^c_+)^m \ket{0;0}$.
This generates e.g. the state with $N$ small and no large excitations $\ket{0;N}=\ket{1\ldots 1}$ and the state with $N$ large and no small excitations $\ket{N;0} = \ket{2\ldots 2}$, and many others in between, see Fig.~\ref{FIG:threels} right panel and App.~\ref{APP:examples} for detailed examples.
In particular, the states in this subspace are non-degenerate and thereby~\cite{breuer2002} their populations obey a Pauli-type rate equation of the form
\begin{align}\label{EQ:paulirate}
    \dot P_{Mm} &= \sum_{M'm'} R_{Mm,M'm'} P_{M'm'}\nn
    &\qquad- \sum_{M'm'} R_{M'm',Mm} P_{Mm}\,,
\end{align}
where $P_{Mm}\equiv\bra{M;m} \rho \ket{M;m}$ and the transition rate from $\ket{M';m'}$ to $\ket{M;m}$ is given by $R_{Mm,M'm'}=\sum_\nu \Big[\gamma_\nu(+\Omega_\nu) \abs{\bra{M;m} \hat{J}^\nu_-\ket{M';m'}}^2+\gamma_\nu(-\Omega_\nu) \abs{\bra{M;m} \hat{J}^\nu_+\ket{M';m'}}^2\Big] \ge 0$.
This rate equation can be directly obtained by evaluating~\eqref{EQ:lindblad} in the fully symmetric basis $\ket{M;m}$, see App.~\ref{APP:pauli} for details.
As Pauli rate equations can be obtained from microscopically derived LGKS equations in special cases (they always result for non-degenerate $\hat{H}_S$, here applied to isolated subspaces), they also obey their favorable properties}, manifest e.g. in the fact that the rates for every reservoir respect local detailed balance~\cite{schnakenberg1976a}.
To obtain the coefficients in Eq.~\eqref{EQ:paulirate}, we require the action of the collective ladder operators in the symmetric subspace, compare Fig.~\ref{FIG:threels} right panel.
They can be evaluated by representing the symmetric subspace with two bosonic modes by means of a generalized Holstein-Primakoff transform, see App.~\ref{APP:bosonization}, which yields
\begin{align}\label{EQ:ladder_coefficients}
\hat{J}^h_+ \ket{M;m} &= \sqrt{(N-M-m)(M+1)} \ket{M+1;m}\,,\nn
\hat{J}^c_+ \ket{M;m} &= \sqrt{(N-M-m)(m+1)} \ket{M;m+1}\,,\nn
\hat{J}^w_+ \ket{M;m} &= \sqrt{(M+1) m} \ket{M+1;m-1}\,,\nn
\hat{J}^h_- \ket{M;m} &= \sqrt{(N-M-m+1)M} \ket{M-1;m}\,,\nn
\hat{J}^c_- \ket{M;m} &= \sqrt{(N-M-m+1)m} \ket{M;m-1}\,,\nn
\hat{J}^w_- \ket{M;m} &= \sqrt{M(m+1)} \ket{M-1;m+1}\,.
\end{align}
Keeping $m=0$ and $M=j_z+N/2$ for $\hat{J}^h_\pm$, 
or keeping $M=0$ and $m=j_z+N/2$ for $\hat{J}^c_\pm$, 
or keeping $M+m=N$ and $M-m=2 j_z$ for $\hat{J}^w_\pm$, 
we see that the transitions along the red or blue or green triangle facets in Fig.~\ref{FIG:threels} right panel, respectively, precisely reproduce the usual $su(2)$ Clebsch-Gordan coefficients of the Dicke states with $j=N/2$.
Accordingly, our model also includes the Dicke superradiant relaxation of two-level systems~\cite{dicke1954a,agarwal1973a,gross1982a} if we couple only to one reservoir.
These Clebsch-Gordan coefficients are largest when $j_z=0$ (i.e., for one green/red/blue bath, they become maximal in the middle of the respective green/red/blue triangle facet in Fig.~\ref{FIG:threels} right panel).
Accordingly, triangles where all coefficients are large have $m\approx M \approx N/3$.
In Fig.~\ref{FIG:threels} right panel, such a central cycle is marked (shaded), and our main findings are based on the properties of these most productive cycles.

Finally, we also use a coarse-grained rate equation~\cite{esposito2012b} valid for infinite temperatures of the work reservoir $n_w(\Delta-\delta)\to\infty$ (see App.~\ref{APP:coarsegraining} for details)
\begin{align}\label{EQ:cgrate}
    \dot Q_n &= \sum_{n'} R_{nn'}^{\rm cg} Q_{n'} - \sum_{n'} R_{n'n}^{\rm cg} Q_n\,,\nn
    Q_n &= \sum_{M,m} P_{Mm} \delta_{M+m,n}\,,
\end{align}
where the mesostate probabilities $Q_n$ are occupations of states with the total number of $n$ (small and/or large) excitations and $\delta_{M+m,n}$ denotes a Kronecker symbol.
This reduction is possible because in this limit, the transitions along the green lines in Fig.~\ref{FIG:threels} right panel become dominant, such that all populations connected by green transitions become identical.
The non-vanishing coarse-grained transition rates then become
\begin{align}\label{EQ:cgrate_rates}
R_{n,n+1}^{\rm cg}&=[\Gamma_c(1+n_c)+\Gamma_h(1+n_h)] \frac{(n+1)(N-n)}{2}\,,\nn
R_{n,n-1}^{\rm cg}&=[\Gamma_c n_c + \Gamma_h n_h] \frac{(n+1)(N+1-n)}{2}
\end{align}
with $\Gamma_\nu \equiv \Gamma_\nu(+\Omega_\nu)$ and $n_\nu\equiv n_\nu(+\Omega_\nu)$ and $0 \le n \le N-1$ in the first and $1 \le n \le N$ in the second line, respectively.
The coarse-grained rate equation provides a tremendous reduction of complexity by mapping our system for $n_w\to\infty$ to a tri-diagonal rate equation, and we provide analytic solutions for the cooling current in App.~\ref{APP:coarsegraining}.

We are dominantly aiming at the steady-state solutions to Eqns.~\eqref{EQ:redfieldmain},~\eqref{EQ:lindblad},~\eqref{EQ:paulirate}, and~\eqref{EQ:cgrate} in non-equilibrium scenarios.
While it is straightforward to evaluate the energy currents $I_{E,S}^\nu(t)$ {\em entering the system} from balances of the system energy
$\frac{d}{dt} \expval{\hat{H}_S} = \sum_\nu I_{E,S}^\nu(t)$, we introduce also microscopically derived counting fields in App.~\ref{APP:derivation}, from which we demonstrate how to obtain the energy currents {\em leaving the reservoirs}
$I_E^\nu = - \frac{d}{dt} \expval{\hat{H}_B^\nu}$ and their fluctuations 
$S_{I_E^\nu} = \frac{d}{dt} \left[\expval{(\hat{H}_B^\nu)^2}-\expval{\hat{H}_B^\nu}^2\right]$
in App.~\ref{APP:fcs}.


\section{Results}\label{SEC:results}

In the collective and steady-state limits of the fully symmetric case (where the stationary solutions of LGKS~\eqref{EQ:lindblad} and Pauli~\eqref{EQ:paulirate} equations are identical), we can establish (see App.~\ref{APP:tightcoupling}), that the stationary currents are tightly coupled 
\begin{align}\label{EQ:tightcoupling}
\bar I_E^w = \frac{\Delta-\delta}{\delta} \bar I_E^c\,,\qquad
\bar I_E^h = -\frac{\Delta}{\delta} \bar I_E^c\,,
\end{align}
and that the $N$-qutrit QAR is subject to the same cooling condition as a single qutrit QAR (see App.~\ref{APP:pauli})
\begin{align}\label{EQ:cooling_condition}
\beta_h \Delta &> \beta_w (\Delta-\delta) + \beta_c \delta\,.
\end{align}
Together with the inherent assumption $\beta_c > \beta_h > \beta_w$, this defines an operational cooling regime.
Due to the tight-coupling, we only quantify the cooling current below.

\begin{figure}
    \centering
    \centerline{\includegraphics[width=0.5\textwidth,clip=true]{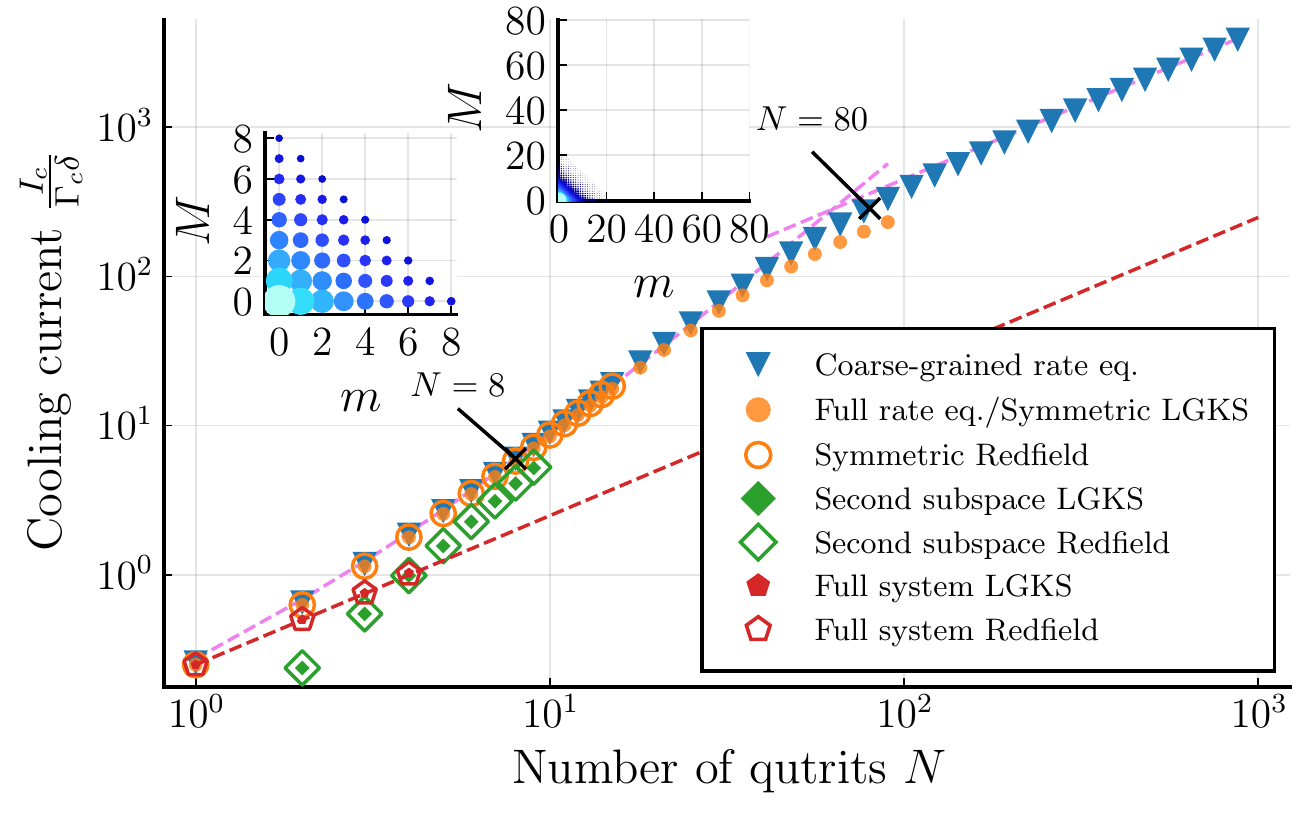}}
    \caption{Double-logarithmic plot of the dimensionless stationary energy current entering the system from the cold reservoir (cooling current) versus the number of qutrits $N$ using different approaches and initial states (symbols, not all orange circles and blue triangles are shown). The dashed \new{violet} curves represent the analytical expressions for $n_w\to\infty$ from Eq.~\eqref{EQ:cg_analytical_current} in the limiting cases $N\ll\bar{n}$ and $N\gg\bar{n}$. \new{The dashed red line represents the linear cooling current scaling generated by $N$ independent qutrits.} Parameters: $\Gamma_c=\Gamma_h=\Gamma_w=0.1\delta,\Delta=10\delta$, $n_c=10$, $n_h=1$ and $n_w=100$. Red pentagons correspond to averages over 100 realizations (resulting error bars are negligibly small) with random-phase couplings like Eq.~\eqref{EQ:coupling} with $h_i^\nu = e^{\ii\varphi_i^\nu}$ and randomly distributed $\varphi_i^\nu\in[-0.1,+0.1]$.
    }
    \label{FIG:currents}
\end{figure}

Our main result however is that, within appropriate regimes, the collective features of our model can support a quadratic scaling of the stationary cooling current with the number of qutrits $N$, see Fig.~\ref{FIG:currents}.
There, we approach the problem with methods of different complexity like the full Redfield treatment, the LGKS treatment, the rate equation treatment, and the coarse-grained rate equation treatment.
We see that all methods coincide for the case $N=1$ (bottom left), where a simple rate equation treatment is sufficient (see App.~\ref{APP:qar1}).
For larger but still small $N$, all methods applicable for a fully collective coupling predict a faster-than-linear growth of the current with the number of qutrits $N$, manifesting the quantum boost of the working fluid.
\new{In App.~\ref{APP:noncollective} we emphasize the quantum nature of this effect.}
This is seen both in LGKS (orange and green filled symbols) and Redfield (orange and green hollow symbols) approaches.
The perfect agreement of LGKS and Redfield approaches (solid versus hollow symbols) demonstrates that in the considered weak-coupling regime, the effect is not just a consequence of the secular approximation. 
Furthermore, for collective couplings we also see that the cooling current in the fully symmetric subspace (orange symbols, generated with ladder operators from state~\eqref{EQ:psirepvac}) is larger than than that originating from a subspace with second largest Casimir operator eigenvalue (green symbols), {which we constructed by acting with collective ladder operators $\hat{J}^\nu_\pm$ on the representative state
\begin{align}\label{EQ:psirep}
    \ket{\Psi_{\rm rep}} &=\frac{1}{\sqrt{N}} \Big[e^{\ii 2\pi 0/N} \ket{0 \ldots 01} + e^{\ii 2\pi 1/N} \ket{0 \ldots 010}\nn
    &\qquad+ \ldots + e^{\ii 2\pi (N-1)/N} \ket{10 \ldots 0}\Big]\,,
\end{align}
until no new orthogonal vectors were found (analogous to the discussion in App.~\ref{APP:examples}).
}

As one may expect for quantum features, the quadratic boost is fragile in some aspects:
First, we observe for large $N$ a crossover to linear scaling (filled orange circles and blue triangles), which we can link to the fact that for fixed temperatures and increasing $N$, the most productive cycles with $m\approx M\approx N/3$ are no longer populated significantly (insets display stationary populations for $N=8$ and $N=80$).
\new{Similiar inhibitions of superradiant behaviour for large $N$ have been observed elsewhere~\cite{vogl2011a, niedenzu2018a}.}
For the limit of an infinitely hot work reservoir (blue triangles), this can even be understood analytically (dashed magenta curves, see App.~\ref{APP:coarsegraining}). 
Second, a more severe restriction appears when we relax the assumption of collective couplings by allowing for random phases in the coupling operators (red symbols).
Then, the  Liouvillian no longer decouples the subspaces of different Casimir operator eigenvalues, and already for weak deviations from the collective limit, the steady-state current no longer scales quadratically.
\new{
In this limit, the steady state is close to a product of single-qutrit states, see App.~\ref{APP:noncollective}, such that 
major parts of the steady state populate less productive subspaces.
}
When the deviations from the collective limit are small, at least two time scales will emerge: A fast one describes the evolution within the subspaces of constant Casimir operator eigenvalue, and the slow one(s) will describe the leakage between the subspaces of different Casimir operator eigenvalues.
Thus, initializing the working fluid in a permutationally symmetric state like $\ket{0;0}$ and operating the device only for a finite time, between these two time scales, may sufficiently populate states near $\ket{N/3;N/3}$ and then still yield a quantum boost.
However, to stabilize the quantum enhancement at steady state and to resist small perturbations, fine-tuned interactions may be required, as exemplified at the end of this section.

To analyze the device from the thermodynamic perspective, we note that LGKS and the derived rate equations are thermodynamically fully consistent. 
At steady state, the first law just implies that the stationary energy currents add to zero $\sum_\nu \bar I_E^\nu=0$ -- see Eq.~\eqref{EQ:tightcoupling}, and the second law implies that the stationary irreversible entropy production rate is non-negative $\bar\sigma_{\ii} = -\sum_\nu \beta_\nu \bar I_E^\nu \ge 0$~\cite{schaller2014}.
As specifically in the symmetric subspace the currents are tightly coupled~\eqref{EQ:tightcoupling}, the coefficient of performance of the device is given by 
$\kappa \equiv \frac{\bar I_E^c}{\bar I_E^w} \Theta(\bar I_E^c)= \frac{\delta}{\Delta-\delta} \Theta(\bar I_E^c)$.
One can check that in the regions of cooling functionality ($\bar I_E^c >0$), the coefficient of performance is bound by its Carnot value via the condition that $\bar\sigma_\ii\ge 0$.
Tighter bounds can be obtained by considering the thermodynamic uncertainty relation~\cite{barato2015a,pietzonka2016a,gingrich2016a}
\begin{align}\label{EQ:tur}
    \bar\sigma_{\ii} \frac{\bar S_{I_E^\nu}}{(\bar I_E^\nu)^2} \ge 2\,.
\end{align}
Using that all currents are tightly coupled, we use it to lower-bound the fluctuations for e.g. the cooling current as
\begin{align}
    \bar S_{I_E^c} \ge \frac{\delta \bar I_E^c}{\beta_h \Delta - \beta_c \delta-\beta_w (\Delta-\delta)}\,,
\end{align}
which proves that the fluctuations inherit the super-extensive scaling of the current \new{(if present)}.
As the eigenvalues of the Pauli rate equation cannot scale faster than $N^2$, we also find that the fluctuations cannot grow faster than $N^2$.
Therefore, the relative fluctuations~\cite{saryal2021a} $\sqrt{\bar S_{I_E^c}}/\bar I_E^c$ 
\new{are also affected by this bound.
Numerically, we find that the fluctuations approximately follow the scaling of the current.
Then, it follows that these are suppressed strongest in the regime of quadratic current and noise scaling.
}
%
Due to the tight-coupling relations~\eqref{EQ:tightcoupling}, we found the relative fluctuations to be alike for all reservoirs~\cite{mohanta2022a}.

\begin{figure}
    \centering
    \includegraphics[width=0.45\textwidth,clip=true]{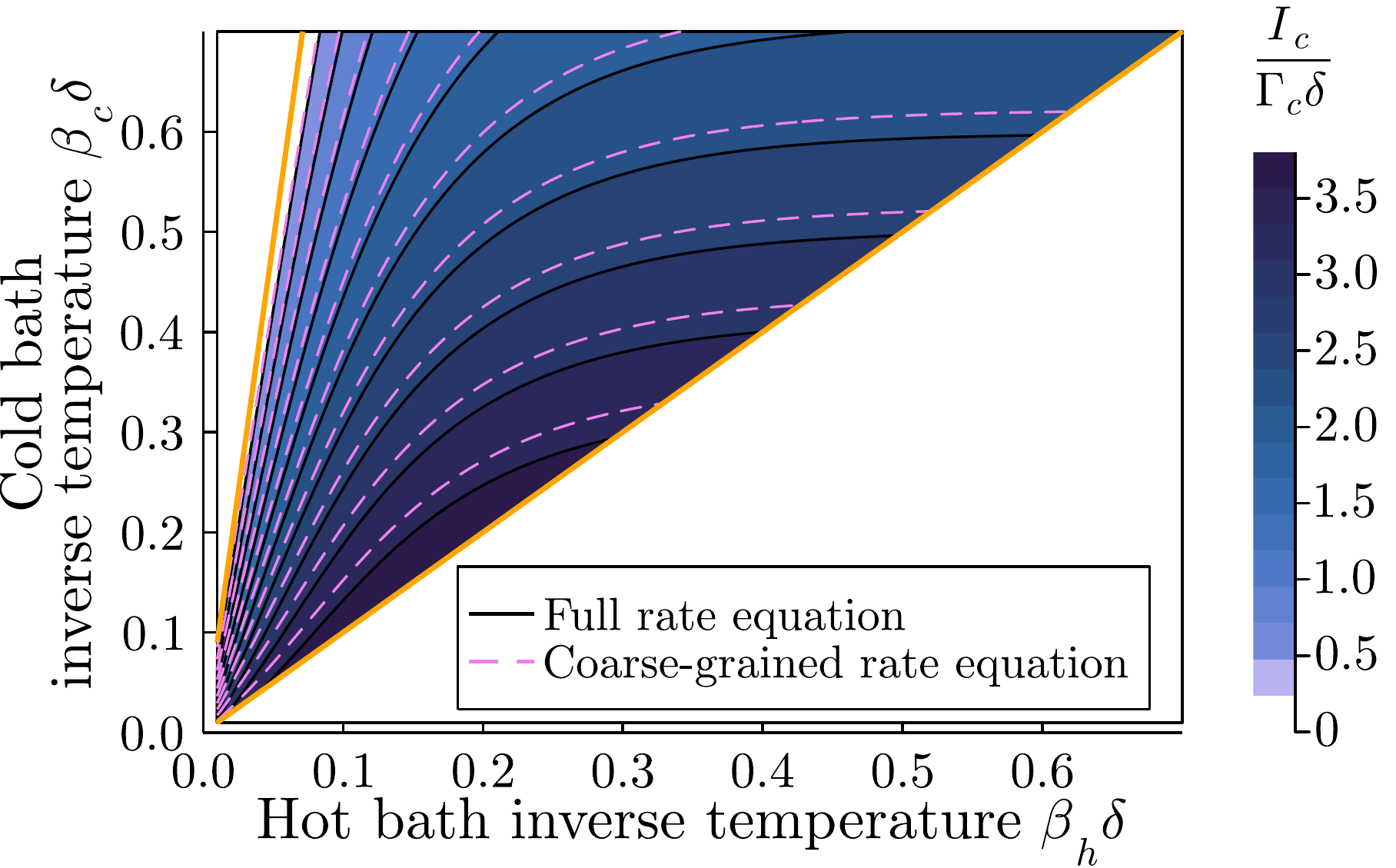}
    \caption{Contour plot of the symmetric subspace cooling current for $N=6$ vs. inverse dimensionless temperatures of hot $\delta \beta_h$ and cold  $\delta \beta_c$ reservoirs. Cooling is achieved in the region bounded by the outer solid lines which represent condition~\eqref{EQ:cooling_condition}. Parameters: $\Gamma_c=\Gamma_h=\Gamma_w$, $\Delta=10\delta$, solid contours: $n_w=100$, dashed contours: $n_w\to\infty$.}
    \label{FIG:contour_colorbar}
\end{figure}

\begin{figure}
    \centering
    \includegraphics[width=0.45\textwidth,clip=true]{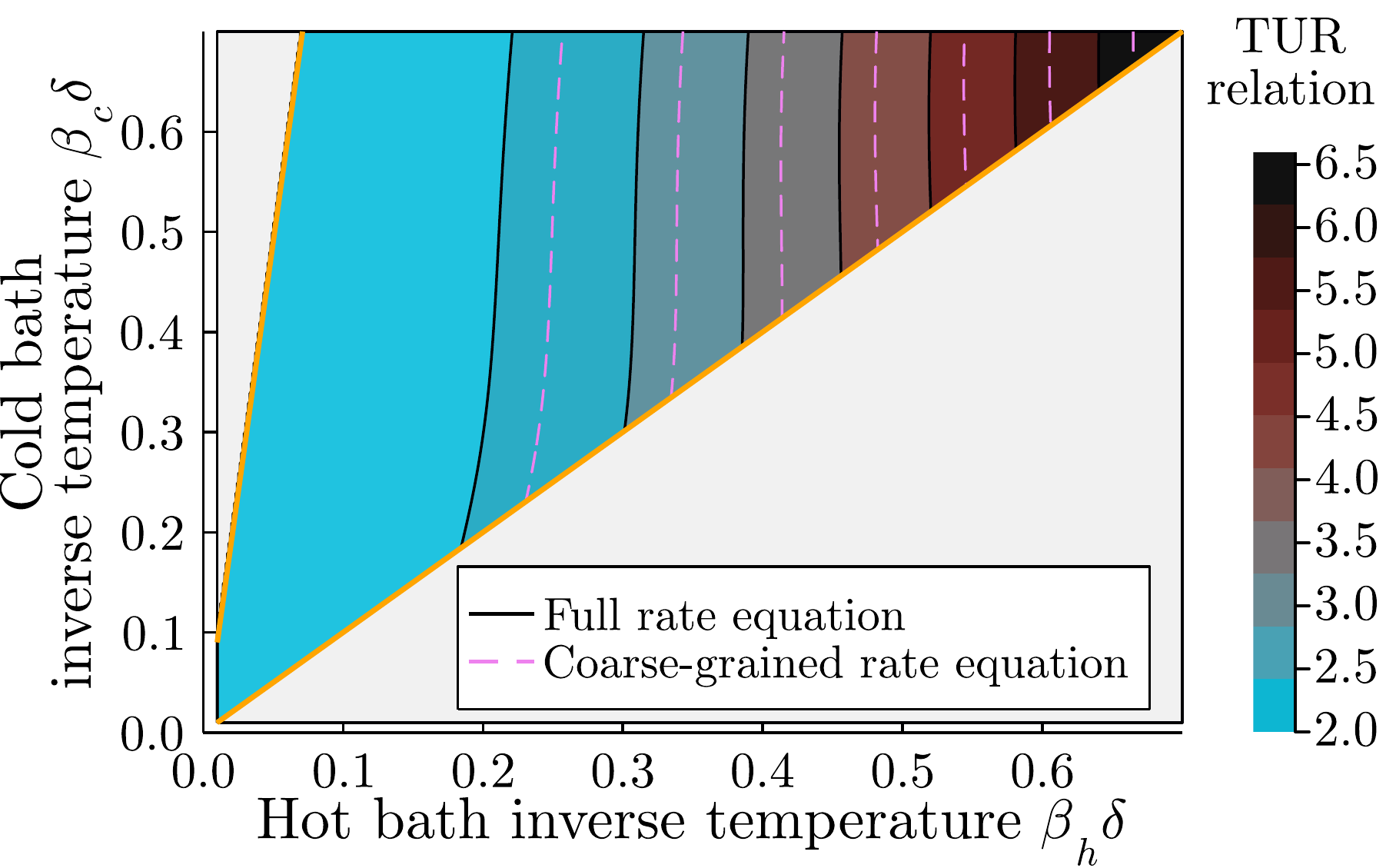}
    \caption{Contour plot analogous to Fig.~\ref{FIG:contour_colorbar} but showing the thermodynamic uncertainty relation~\eqref{EQ:tur} in the cooling region. Parameters as in Fig.~\ref{FIG:contour_colorbar}.
    }
    \label{FIG:tur_plot_gr}
\end{figure}

Thus, if in practice one would like a device with a large cooling power, a look at 
Fig.~\ref{FIG:contour_colorbar} suggests that our device should be operated at $\beta_c \gtrapprox \beta_h$.
Likewise, one would would like to have a device with \new{a small product of relative fluctuations and overall entropy production, as quantified by the uncertainty quotient~\eqref{EQ:tur}.
We find that this quotient is} smallest at the other boundary of the cooling region, where $\beta_c \lessapprox \beta_h \frac{\Delta}{\delta}-\beta_w \frac{\Delta-\delta}{\delta}$, see Fig.~\ref{FIG:tur_plot_gr}.
Thus, our model recovers the usual trade-off between accuracy and performance. 

\begin{figure}
    \centering
    \includegraphics[width=0.45\textwidth,clip=true]{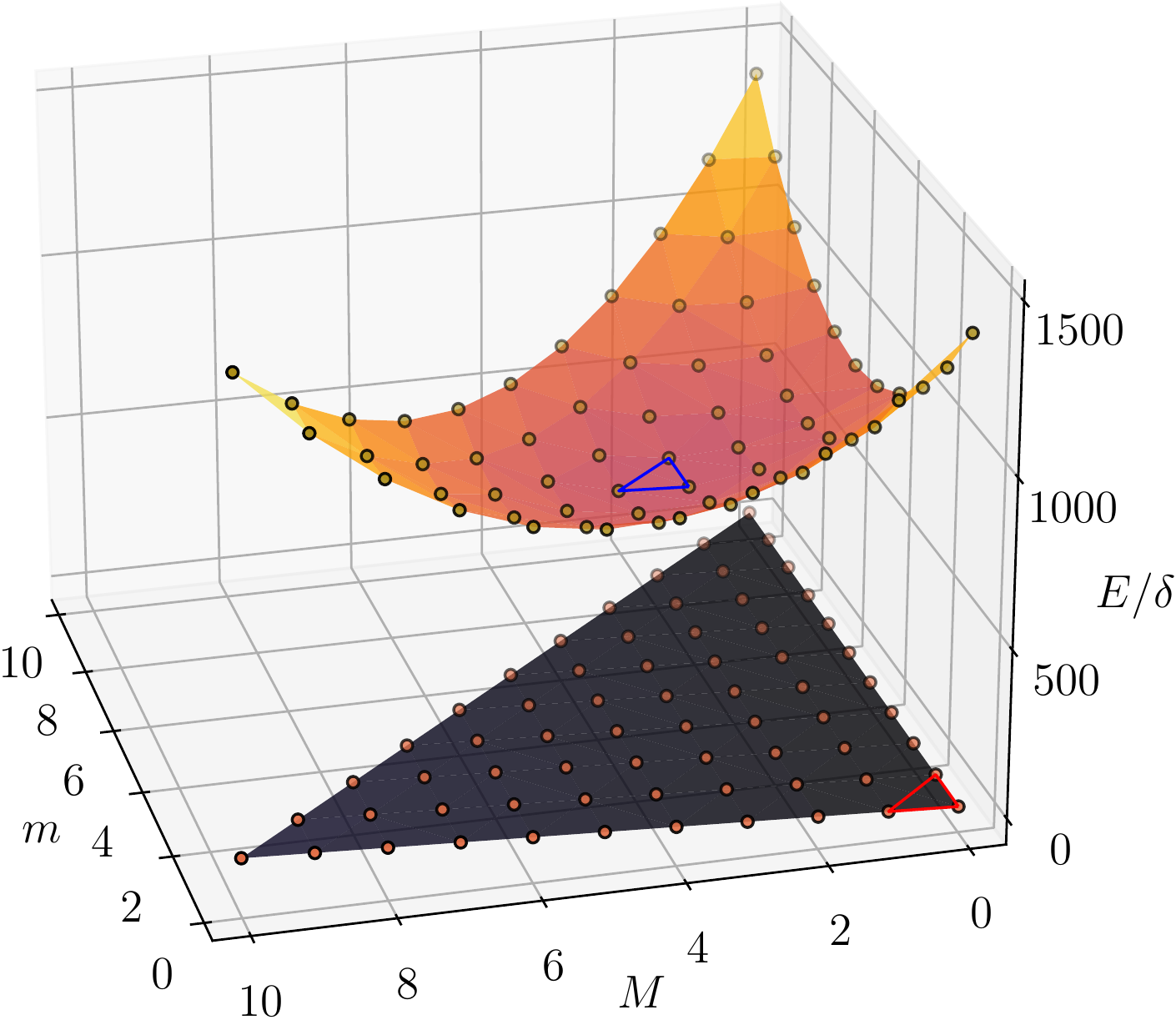}
    \caption{Visualization of $\hat{H}_S + \Delta \hat{H}_S$ for $N=10$ and within the maximally symmetric subspace. The bottom plane depicts the energy landscape in absence of a penalty Hamiltonian, and the energetically most favorable triangle (red) is at lowest excitation numbers. The curved surface demonstrates the minimal penalization of the most productive triangle (blue), the plane defined by it is parallel to the plane without interactions. Parameters: $\Delta=10\delta$, $\alpha_P=20 \delta$ (the curved surface has been shifted upwards for clarity).
    }
    \label{FIG:energy_landscape}
\end{figure}

By adding specifically taylored interactions to the system, we can energetically favor the maximally symmetric subspace and also the cycle with the maximum current.
One example for such an interaction could be
\begin{align}\label{EQ:ham_penalty}
    \Delta \hat{H}_S &= \alpha_C \left[\frac{N(N+3)}{3}-\hat{C}_2\right]+ \alpha_P \Bigg[\left(\frac{N}{3}-\hat{N}_\Delta\right)^2\nn
    &\quad+\left(\frac{N}{3}-\hat{N}_\delta\right)^2
    +\left(\frac{N}{3}-\hat{N}_\Delta\right)\left(\frac{N}{3}-\hat{N}_\delta\right)\Bigg]
\end{align}
with coefficients $\alpha_C>0$ and $\alpha_P>0$ penalizing the deviation from the maximum Casimir sector and the central triangle, respectively (all operators in square brackets are positive semidefinite).
For the case where $N=3k+1$ with integer $k$ (for other configurations one may adapt the penalty Hamiltonian accordingly), the three states of the maximum symmetry sector $\ket{k;k}$, $\ket{k;k+1}$, and $\ket{k+1;k}$ have the same minimal energy penalty, see the blue triangle in Fig.~\ref{FIG:energy_landscape}.
Therefore, it cancels out in the difference of energies, and the corresponding transition rates are the same as without any interaction.
Provided that the energy penalty is strong enough $\beta_\nu \alpha_{C/P} \gg 1$, the other states with larger energy penalties can be omitted from the considerations, and we obtain the current for a single QAR as discussed in Eq.~\eqref{EQ:qarn1p} but with boosted rates 
$\Gamma_\nu\to \Gamma_\nu \left(\frac{(N+2)}{3}\right)^2$, which results from the enhanced Clebsch-Gordan coefficients~\eqref{EQ:ladder_coefficients} in the central cycle.
Accordingly, for $N$ qutrits the cooling current will be boosted 
\begin{align}\label{EQ:current_boost}
    \bar I_{E}^{c,N} \approx \left(\frac{(N+2)}{3}\right)^2 \bar I_E^{c,1}
\end{align}
compared to the current from a single QAR $\bar I_E^{c,1}$. 
As we scale all coupling constants, the fluctuations (and all higher cumulants) would increase likewise, 
such that both $\bar I_E^c \propto N^2$ and $\bar S_{I_E^c} \propto N^2$.
%

\section{Conclusions}\label{SEC:conclusions}

We analyzed an ensemble of $N$ identical qutrits for its suitability as a QAR using different approaches. 
The most striking effect is a quadratic scaling of the output (cooling current) with the number of qutrits for a permutationally symmetric configuration and collective system-reservoir couplings.
This collective phenomenon can be understood analogously to Dicke superradiant decay of two level systems -- here put to use in an engine by collective couplings.
In contrast to quantum computers subjected to decoherence, the performance of the QAR is only reduced to linear (classical) scaling for too large $N$ (changing other parameters may increase the critical size) or for a not-perfectly-collective coupling. 
For non-identical qutrits we would expect a similar breakdown~\cite{gerry2022a}.
Compared to a qubit implementation~\cite{kloc2021a}, we want to summarize some advantages of collective qutrits:
First, the initialization in the proper subspace requires just a permutationally symmetric state (no entangled states like in~\cite{kloc2021a}) as e.g. $\ket{0\ldots 0}$.
Second, to see the speedup, we do not require fine-tuned inter-qutrit interactions (although this can be used to stabilize the setup against perturbations).
Third, in contrast to the interacting qubit system, the (non-interacting) multi-qutrit system only provides three transition frequencies, such that a selective driving appears easier to implement.

On the technical side, we have confirmed the effect with a variety of methods. 
Using a bosonization technique is helpful to compute the transition rates analytically.
While the behaviour of the collective limit can be well understood with analytic approximations and allowed us to treat fairly large systems, treating the non-collective limit requires significant resources. 
We have benchmarked our results with considering also the non-LGKS Redfield master equation. 
We found that LGKS and Redfield results did not differ much in the considered weak-coupling regime.
Beyond stronger system-reservoir coupling strengths, one may also expect larger deviations outside the tight-coupling regime, where the reservoirs no longer drive the individual transitions.

As an outlook, we think it could be interesting to consider laser-driven QARs~\cite{kalaee2021a,kalaee2021b} or investigate the fluctuations of energy currents beyond the tight-coupling regime~\cite{jaseem2022a} or for detuned levels~\cite{holubec2019a} in greater detail.
One may also be tempted to address the strong-coupling regime beyond phenomenologic models~\cite{mu2017a}, e.g. using reaction coordinates~\cite{restrepo2018a}.
This however should be performed while maintaining a lower spectral bound on the global Hamiltonian for all coupling strengths and $N$, in this case we expect modifiations to Eq.~\eqref{EQ:hamgeneral}~\cite{nataf2010a}.

\acknowledgments

D. K. acknowledges support by the HZDR summer student program. 
G. S. has been supported by the Helmholtz high potential program and the DFG (CRC 1242, project-ID 278162697).


\bibliographystyle{unsrt}
\bibliography{references}

\appendix


\section{A single three-level QAR}\label{APP:qar1}

For the case of a single qutrit $N=1$ one may derive a rate equation
$\dot{\vec{P}} = R \vec{P}$ for the occupation probabilities $\vec{P}=(P_0, P_1, P_2)^{\rm T}$
of these states (Redfield, LGKS, and Pauli rate equations then yield the same dynamics for these probabilities).

For the normal configuration shown in Fig.~\ref{FIG:threels} bottom left panel, where the cold reservoir drives transitions between the lowest two states $\ket{0}\leftrightarrow\ket{1}$, the hot between lowest and highest $\ket{0}\leftrightarrow\ket{2}$, and the work reservoir transitions between the two excited states $\ket{1}\leftrightarrow\ket{2}$,
the rate matrix becomes
\begin{align}\label{EQ:qarn1p}
    R^+ = 
    \left(\begin{array}{ccc}
    R_{00}^+ & \Gamma_c(1+n_c) & \Gamma_h (1+n_h)\\
    \Gamma_c n_c &  R_{11}^+ & \Gamma_w(1+n_w)\\
    \Gamma_h n_h & \Gamma_w n_w & R_{22}^+
    \end{array}\right)\,,
\end{align}
where $n_\nu \equiv n_\nu(\Omega_\nu)$ and the negative diagonal elements are fixed by demanding vanishing of the column sums (this implements overall probability conservation).
In the rate matrix, we can clearly distinguish the contributions of the individual reservoirs.
From this, the energy current entering from, for example, the cold reservoir can be obtained by standard techniques:
One may compute the current via $I_E^c(t) = \sum_{ij} (E_i-E_j) R_{ij,c}^+ P_j(t)$, where $R_c^+=R^+|_{\Gamma_h=\Gamma_w=0}$ is the cold reservoir rate matrix.
Alternatively, 
one may introduce energy counting fields $R^+\to R^+(\chi)$ by replacing $\Gamma_c(1+n_c)\to\Gamma_c(1+n_c) e^{-\ii\delta\chi}$ and $\Gamma_c n_c\to \Gamma_c n_c e^{+\ii\delta\chi}$ in the off-diagonal elements of the rate matrix and compute the current via $I_E^c(t) = -\ii (1,1,1)(\partial_\chi R^+(\chi))|_{\chi=0} \vec{P}(t)$, see also App.~\ref{APP:fcs}.
With this formalism, it is possible to compute also higher cumulants of the distribution of energy transfers~\cite{segal2018a}.
In the limit of $n_w\to\infty$, the long-term (stationary) current reduces to 
\begin{align}\label{EQ:current_qar1}
\bar I_{E+}^{c,\infty} = \frac{\Gamma_c \Gamma_h \delta (n_c-n_h)}{\Gamma_c (1+3 n_c) + \Gamma_h(1+3 n_h)}\,.
\end{align}

For the case where cold and work reservoir are exchanged $E_0=0$, $E_1=\Delta-\delta$, $E_2=\Delta$ depicted in Fig.~\ref{FIG:threels} top left panel,
the rate matrix is given by
\begin{align}
    R^-=\left(\begin{array}{ccc}
    R_{00}^- & \Gamma_w(1+n_w) & \Gamma_h (1+n_h)\\
    \Gamma_w n_w & R_{11}^- & \Gamma_c(1+n_c)\\
    \Gamma_h n_h & \Gamma_c n_c & R_{22}^-
    \end{array}\right)\,,
\end{align}
and an analogous calculation leads in the case of $n_w\to\infty$ to a slightly different cooling current
\begin{align}
\bar I_{E-}^{c,\infty} = \frac{\Gamma_c \Gamma_h \delta (n_c-n_h)}{\Gamma_c (2+3 n_c) + \Gamma_h(2+3 n_h)}\,.
\end{align}

Both currents have the same cooling condition
\begin{align}\label{EQ:coolingcond0}
n_c > n_h \qquad\Leftrightarrow\qquad
\beta_c < \beta_h \Delta/\delta\,,
\end{align}
which -- together with $\beta_c>\beta_h$ -- 
defines an operational window for cooling the coldest reservoir.
This is a sufficient and necessary condition for $n_w \to \infty$ and $N=1$.

For finite $n_w$, the calculations are a bit lengthier but fully analogous. 
One obtains for both configurations the sufficient and necessary cooling condition
\begin{align}\label{EQ:coolingcond1}
    \beta_w (\Delta-\delta) + \beta_c \delta < \beta_h \Delta\,,
\end{align}
which reduces to the previous condition as $\beta_w\to 0$.
Together with the underlying assumption $\beta_c > \beta_h$, this defines the bounds on the window of cooling also for larger $N$ (compare Fig.~\ref{FIG:contour_colorbar} in the main text).


\section{Collective spin representation}\label{APP:gellmann} 

For a single qutrit, we can express the system Hamiltonian as
\begin{align}
\hat{H}_S^1 &= (\Delta-\delta) \frac{\hat{\lambda}^3}{2} + (\Delta+\delta) \frac{\hat{\lambda}^8}{2\sqrt{3}} + (\Delta+\delta) \frac{\f{1}}{3}\,,
\end{align}
where the $\hat{\lambda}^{3/8}$ are the two diagonal Gell-Mann matrices.

With using the collective qutrit operators~\eqref{EQ:collective}, we can write the system Hamiltonian as
\begin{align}
\hat{H}_S = (\Delta-\delta) \hat{J}_3 + \frac{1}{\sqrt{3}} (\Delta+\delta) \hat{J}_8 + \frac{N}{3} (\Delta+\delta) \f{1}\,.
\end{align}

Furthermore, we define collective raising and lowering operators
\begin{align}
\hat{J}^h_\pm &= \sum_{i=1}^N \hat{\lambda}^h_{\pm,i} = \hat{J}_4 \pm \ii \hat{J}_5\,,\nn
\hat{J}^c_\pm &= \sum_{i=1}^N \hat{\lambda}^c_{\pm,i} = \hat{J}_6 \pm \ii \hat{J}_7\,,\nn
\hat{J}^w_\pm &= \sum_{i=1}^N \hat{\lambda}^w_{\pm,i} = \hat{J}_1 \pm \ii \hat{J}_2\,,
\end{align}
for which we find an interaction picture dynamics analogous to Eq.~\eqref{EQ:intpic}.

In the collective limit (where $\hat{S}^\nu \to \hat{J}^\nu = \hat{J}^\nu_+ + \hat{J}^ \nu_-$), 
it follows that the quadratic and cubic Casimir operators of the $su(3)$ 
\begin{align}
\hat{C}_2 &= \sum_{\alpha=1}^8 \hat{J}_\alpha^2\,,\nn
\hat{C}_3 &=\sum_{\alpha\beta\gamma} \trace{\left\{\hat{J}_\alpha, \hat{J}_\beta\right\} \hat{J}_\gamma} \hat{J}_\alpha \hat{J}_\beta \hat{J}_\gamma
\end{align}
will be automatically conserved, to all orders in the system-reservoir interaction Hamiltonian.
Depending on the initial condition, this strongly reduces the Hilbert space dimension that needs to be treated explicitly. 
For example, assuming that our system is prepared e.g. in the collective ground state~\eqref{EQ:psirepvac}, 
we can constrain ourselves to the subspace of $(N+1)(N+2)/2$ permutationally symmetric states, which is significantly less demanding than treating $3^N$ basis states~\cite{gegg2017a,silva2022a}.

We label these states analogously to the maximum angular momentum Dicke states (known from collective two-level systems) by $\ket{M;m}$ with
$0 \le M$ large and $0 \le m$ small excitations such that $0 \le M+m \le N$, specific examples are provided in App.~\ref{APP:examples}.
These states are eigenstates of $\hat{N}_\Delta$ and $\hat{N}_\delta$ (or alternatively $\hat{J}_3$ and $\hat{J}_8$) and the Casimir operators, in particular we have
\begin{align}
\hat{N}_\Delta \ket{M;m} &= M \ket{M;m}\,,\qquad
\hat{N}_\delta \ket{M;m} = m \ket{M;m}\,,\nn
\hat{J}_3 \ket{M;m} &= \left[\frac{M-m}{2}\right] \ket{M;m}\,,\nn
\hat{J}_8 \ket{M;m} &= \left[\frac{\sqrt{3}}{2} (M+m)-\frac{N}{\sqrt{3}}\right] \ket{M;m}\,,\nn
\hat{C}_2 \ket{M;m} &= \frac{N(N+3)}{3} \ket{M;m}\,,\nn
\hat{C}_3 \ket{M;m} &= \frac{N(N+3)(2N+3)}{18} \ket{M;m}\,.
\end{align}
Clearly, they are also eigenstates of the system Hamiltonian $\hat{H}_S \ket{M;m} = \left(M \Delta + m \delta\right) \ket{M;m}$.


\section{Example states for finite $N$}\label{APP:examples}

In total, the Hilbert space dimension for $N$ qutrits is $D=3^N$.
This can be decomposed by counting the number of states with $M$ large and $m$ small excitations
\begin{align}
N_{Nm} =  \frac{N!}{(N-M-m)! M! m!}\,,
\end{align}
and indeed we have $3^N = \sum\limits_{M=0}^N \sum\limits_{m=0}^{N-M} N_{Mm}$.
However, the subspace of completely symmetric states under permutations has only 
\begin{align}
N_{\rm symm} = \frac{(N+1)(N+2)}{2}
\end{align}
elements, which are closed under the action of $\hat{J}^\nu_{\pm}$.
This subspace is characterized by the largest Casimir operator eigenvalues.
Therefore, many states belong to subspaces which are not symmetric under permutations, see below for $N=2$ and $N=3$, where we take the convention of labeling the local eigenstates of the single-qutrit Gell-Mann matrices $\hat\lambda^3$ and $\hat\lambda^8$ as $\ket{0}$, $\ket{1}$, and $\ket{2}$.

\subsection{Example states for $N=2$}

For $N=2$, we have 6 out of the $9=3^2$ states in total belonging to the completely symmetric subspace 
\begin{align}
\ket{0;0} &= \ket{00}\,,\qquad
\ket{0;2} = \ket{11}\,,\qquad
\ket{2;0} = \ket{22}\,,\nn
\ket{0;1} &= \frac{1}{\sqrt{2}} \left[\ket{01}+\ket{10}\right]\,,\qquad
\ket{1;0} = \frac{1}{\sqrt{2}} \left[\ket{02}+\ket{20}\right]\,,\nn
\ket{1;1} &= \frac{1}{\sqrt{2}} \left[\ket{21}+\ket{12}\right]\,,
\end{align}
and due to the complete permutational symmetry of the $\hat{J}^\nu_{\pm}$ the evolution of these six states is closed under collective couplings.
These states are eigenstates of $\hat{C}_2$ and $\hat{C}_3$ with eigenvalues $10/3$ and $35/9$, respectively.
Trivially, cyclic permutations leave these states invariant.
The two-boson representation from App.~\ref{APP:bosonization} with $N_a=2$ and $N_b=0$ suffices to represent this subspace with $M$ and $m$ representing the eigenvalues of $\hat{N}_\Delta$ and $\hat{N}_\delta$, respectively.
One can check the action of the ladder operators~\eqref{EQ:ladder_coefficients} among them.

Additionally, we have the three antisymmetric states
\begin{align}
\ket{\Psi_7} &= \frac{1}{\sqrt{2}}\left[\ket{01}-\ket{10}\right]\,,\nn
\ket{\Psi_8}\ &= \frac{1}{\sqrt{2}} \left[\ket{21}-\ket{12}\right] = \hat{J}^h_+ \ket{\Psi_7}\,,\nn
\ket{\Psi_9} &= \frac{1}{\sqrt{2}} \left[\ket{02}-\ket{20}\right] = \hat{J}^w_+ \ket{\Psi_7}\,,
\end{align}
whose evolution is also closed under collective couplings.
These states are eigenstates of the Casimir operators $\hat{C}_2$ and $\hat{C}_3$ with eigenvalues $4/3$ and $-10/9$, respectively.
Cyclic permutations of the qutrits equip these states with a phase factor of $-1=e^{\ii 2\pi/2}$.
In the four-boson representation $\ket{M,m,Q,q}$ with $N_a=0$ and $N_b=1$ we would identify them with the states $\ket{\Psi_7}=\ket{0,0,0,0}$, 
$\ket{\Psi_8} = \ket{0,0,1,0}$, and $\ket{\Psi_9} = -\ket{0,0,0,1}$, compare Eq.~\eqref{EQ:4mode_holstein_primakoff}.

\subsection{Example states for $N=3$}

For $N=3$, we can build the completely symmetric subspace by starting from the representative state $\ket{\Psi_1} = \ket{0;0}=\ket{000}$ (all atoms in the ground state), and we can generate e.g. 
\begin{align}
\ket{0;1} &= \frac{1}{\sqrt{3}} \left[\ket{001}+\ket{010}+\ket{100}\right]\,,\nn
&= \frac{1}{\sqrt{3}} \hat{J}^c_+ \ket{0;0}\,,\nn
\ket{1;1} &=\frac{1}{\sqrt{6}}\big[\ket{201}+\ket{012}+\ket{120}\nn
&\qquad+\ket{021}+\ket{210}+\ket{102}\big]\nn
&= \frac{1}{\sqrt{2}} \hat{J}^h_+ \ket{0;1}\,,\nn
\ket{1;2} &= \frac{1}{\sqrt{3}}\left[\ket{211}+\ket{112}+\ket{121}\right]\nn
&= \frac{1}{\sqrt{2}}  \hat{J}^c_+ \ket{1;1}
\end{align}
and further states.
That way, we obtain $(3+1)(3+2)/2=10$ completely symmetric states, and we can check the coefficients between them given in Eq.~\eqref{EQ:ladder_coefficients}.
These are eigenstates of the Casimir operators $\hat{C}_2$ and $\hat{C}_3$ with eigenvalues $6$ and $9$, respectively, and can be represented by two bosonic modes $\ket{M;m}$ with $0 \le M+m \le N_a=N=3$ and $N_b=Q=q=0$.
Arbitrary permutations leave these states invariant (in other words, they equip them with a phase of $1$).

In addition to these, we have states with the same number of excitations as $\ket{0;1}$ but which are orthogonal to each other and also to $\ket{0;1}$.
From these, we can also build further sets by acting with the ladder operators. 
For example, starting from $\ket{\Psi_{11}}$ that is orthogonal to $\ket{0;1}$, we get in total 7 states
\begin{align}
\ket{\Psi_{11}} &= \frac{1}{\sqrt{3}}\left[\ket{001}+e^{\ii 2\pi/3} \ket{010}+e^{\ii 4\pi/3} \ket{100}\right]\,,\nn
\hat{J}^c_+ \ket{\Psi_{11}} &= \frac{e^{\ii\pi}}{\sqrt{3}}\left[\ket{110} + e^{\ii \pi 2/3} \ket{101}+e^{\ii\pi 4/3} \ket{011}\right]\,,\nn
\hat{J}^h_+ \ket{\Psi_{11}} &= \frac{\sqrt{2}}{\sqrt{6}} \Big[\ket{201}+e^{\ii 2\pi/3} \ket{012}+ e^{\ii 4\pi/3} \ket{120}\nn
&\qquad+\ket{021}+e^{\ii 2\pi/3} \ket{210}+e^{\ii 4\pi/3} \ket{102}\Big]\,,\nn
\hat{J}^w_+ \ket{\Psi_{11}} &= \frac{1}{\sqrt{3}}\left[\ket{002}+e^{\ii 2\pi/3} \ket{020}+e^{\ii 4\pi/3} \ket{200}\right]\,,\nn
\hat{J}^h_+ \hat{J}^c_+ \ket{\Psi_{11}} &= \frac{e^{\ii\pi} }{\sqrt{3}}\left[\ket{112} + e^{\pi \ii 2/3} \ket{121} + e^{\ii\pi 4/3} \ket{211}\right]\,,\nn
\hat{J}^h_+ \hat{J}^w_+ \ket{\Psi_{11}} &= \frac{e^{\ii\pi} }{\sqrt{3}}\left[\ket{220} + e^{\ii \pi 2/3} \ket{202} + e^{\ii\pi 4/3} \ket{022} \right]\,,\nn
\hat{J}^h_+ \hat{J}^h_+ \ket{\Psi_{11}} &= \frac{2}{\sqrt{3}}\left[\ket{221}+e^{\ii 2\pi/3} \ket{212} + e^{\ii 4\pi/3} \ket{122}\right]\,.
\end{align}
Cyclic permutations equip these with a phase of $e^{\ii 2\pi/3}$.
Further states can be generated when starting from $\ket{\Psi_{18}}$ that is orthogonal to $\ket{\Psi_{11}}$ and $\ket{0;1}$
\begin{align}
\ket{\Psi_{18}} &= \frac{1}{\sqrt{3}}\left[\ket{001}+e^{\ii 4\pi/3} \ket{010}+e^{\ii \pi 2/3} \ket{100}\right]\,,\nn
\hat{J}^c_+ \ket{\Psi_{18}} &= \frac{e^{\ii\pi}}{\sqrt{3}} \left(\ket{110} + e^{\ii \pi 4/3} \ket{101} + e^{+\ii\pi 2/3}\ket{011}\right]\,,\nn
\hat{J}^h_+ \ket{\Psi_{18}} &= \frac{\sqrt{2}}{\sqrt{6}} \Big[\ket{201}+e^{\ii\pi 4/3} \ket{012}+ e^{\ii\pi 2/3} \ket{120}\nn
&\qquad+\ket{021}+e^{\ii\pi 4/3} \ket{210}+ e^{\ii\pi 2/3} \ket{102}\Big]\,,\nn
\hat{J}^w_+ \ket{\Psi_{18}} &= \frac{1}{\sqrt{3}} \left[\ket{002}+e^{\ii\pi 4/3} \ket{020} + e^{\ii\pi 2/3} \ket{200}\right]\,,\nn
\hat{J}^h_+ \hat{J}^c_+ \ket{\Psi_{18}} &= \frac{e^{\ii\pi}}{\sqrt{3}} \left[\ket{112} + e^{\pi\ii 4/3} \ket{121} + e^{+\pi\ii 2/3}\ket{211}\right]\,,\nn
\hat{J}^h_+ \hat{J}^h_+ \ket{\Psi_{18}} &= \frac{2}{\sqrt{3}} \left[\ket{221} + e^{\ii\pi 4/3} \ket{212} + e^{\ii\pi 2/3} \ket{122}\right]\,,\nn
\hat{J}^h_+ \hat{J}^w_+ \ket{\Psi_{18}} &= \frac{e^{\ii\pi}}{\sqrt{3}}\left[\ket{220}  + e^{\pi\ii 4/3} \ket{202} + e^{\pi\ii 2/3} \ket{022}\right]\,.
\end{align}
Under cyclic permutations, these get a phase of $e^{\ii 4\pi/3}$.
We have two additional states with exactly one small and one large excitation that are orthogonal to $\ket{1;1}$, $\hat{J}^h_+ \ket{\Psi_{11}}$, $\hat{J}^h_+ \ket{\Psi_{18}}$ and to each other and having a similar behavior under cyclic permutations.
These are
\begin{align}
\ket{\Psi_{25}} &=\frac{1}{\sqrt{6}} \Big[\ket{201}+e^{\ii\pi 2/3}\ket{012}+e^{\ii \pi 4/3} \ket{120}\nn
&\qquad-\ket{021}-e^{\ii\pi 2/3}\ket{210}-e^{\ii \pi 4/3}\ket{102}\Big]\,,\nn
\ket{\Psi_{26}} &= \frac{1}{\sqrt{6}} \Big[\ket{201}+e^{\ii\pi 4/3} \ket{012}+ e^{\ii\pi 2/3} \ket{120}\nn
&\qquad-\ket{021}-e^{\ii\pi 4/3} \ket{210}- e^{\ii\pi 2/3} \ket{102}\Big]\,.
\end{align}
The first one gets a phase $e^{\ii 2\pi/3}$ under cyclic permutations, the second the phase $e^{\ii 4\pi/3}$.
Additionally, one can see that the states $\hat{J}^h_+\ket{\Psi_{11}}$, $\hat{J}^h_+\ket{\Psi_{18}}$ are symmetric under the state-conditional permutation $0\leftrightarrow 2$, whereas
the states $\ket{\Psi_{25}}$ and $\ket{\Psi_{26}}$ are antisymmetric.
One can get to $\ket{\Psi_{25}}$ e.g. via $\hat{J}^w_+ \hat{J}^c_+ \ket{\Psi_{11}}$ and subsequent orthonormalization.
Analogously, $\ket{\Psi_{26}}$ can be reached by $\hat{J}^w_+ \hat{J}^c_+ \ket{\Psi_{18}}$ and subsequent orthonormalization.
Thus, the $16=2\times 8$ states $\ket{\Psi_{11}}, \ldots ,\ket{\Psi_{26}}$ are closed under the action of a collective reservoir, they have eigenvalues of Casimir operators $\hat{C}_2$ and $\hat{C}_3$ of $3$ and $0$, respectively.
In the bosonic four-mode representation $\ket{M,m,Q,q}$, the $8$ states for this Casimir subspace and given behaviour under cyclic permutations can be constructed from the subspaces with $N_a=1$ and $N_b=1$.
Formally, the bosonic four-mode representation yields via $(M,m),(Q,q)\in\{(0,0),(0,1),(1,0)\}$ in total $9$ (6 non-degenerate and 3 degenerate) possible states, but one superposition of the three degenerate states $\ket{0,0,1,0}$, $\ket{0,1,0,1}$, and $\ket{1,0,0,0}$ has a different Casimir operator eigenvalue and can be decoupled.

Finally, the last state is
\begin{align}
\ket{\Psi_{27}} &=\frac{1}{\sqrt{6}}\Big[\ket{012}+\ket{120}+\ket{201}\nn
&\qquad-\ket{021}-\ket{210}-\ket{102}\Big]\,.
\end{align}
It is annihilated by $\hat{J}^\nu_{\pm}$, has the smallest eigenvalue $0$ of both Casimir operators $\hat{C}_i$, 
and is also inert under cyclic permutations.
It is fully antisymmetric under the exchange of any two qutrits, and corresponds in the bosonic four-mode representation to the state $\ket{0,0,0,0}$ with $N_a=0$ and $N_b=0$.

We note that in the collective limit, states like $\ket{\Psi_{27}}$ remain fully inert, they are dark states~\cite{hayn2011a,orioli2022a}.
From the combinatorics of states, we may expect dark states with $m=M=N/3$ for any number $N$ divisible by 3.
These dark states will then have minimum, i.e., vanishing, Casimir operator eigenvalue $\expval{\hat{C}_2}=0$.
In particular, since already the case $N=6$ hosts more than one dark state, superpositions of these could be used to form e.g. a decoherence-protected logical qubit in presence of all possible collective system-reservoir interactions.

\subsection{States for $N=4$}

\begin{figure}
\includegraphics[width=0.49\textwidth,clip=true]{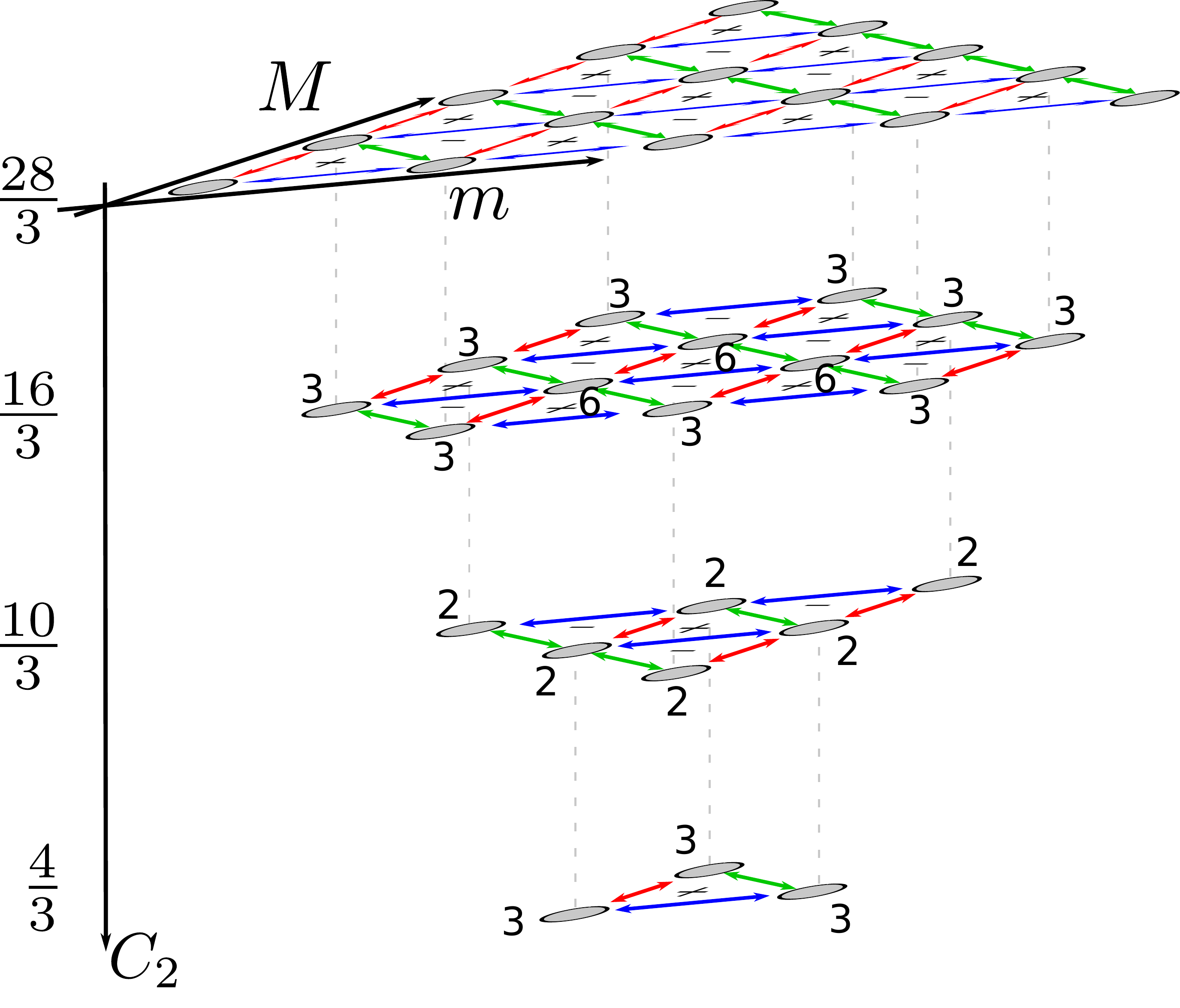}
\caption{\label{FIG:threels4tot}
Map of all $3^4=81$ collective basis states for $N=4$ versus small and large numbers of excitations (horizontal directions) and versus Casimir operator $\hat{C}_2$ eigenvalue.
The top layer corresponds to the $15$ non-degenerate states from the completely permutationally symmetric subspace with largest Casimir operator eigenvalue as shown in Fig.~\ref{FIG:threels} right panel.
It evolves independently from the other states for fully collective transitions.
The other levels are grouped according to decreasing Casimir operator eigenvalue, with numbers \new{close to} symbols indicating the degeneracy.
}
\end{figure}

For $N=4$, we may consider Fig.~\ref{FIG:threels} right panel as the largest (top) layer of all states that can be grouped in a pyramid-like structure as shown in Fig.~\ref{FIG:threels4tot}.


\section{Derivation of the evolution equations}\label{APP:derivation}

\subsection{Redfield master equation}\label{APP:redfield}

\new{
For multiple weakly-coupled reservoirs, the dissipators act additively, such that for a single-coupling operator per reservoir
$\hat{H}_I^\nu = \hat{S}^\nu \otimes \hat{B}^\nu$, the Redfield-II master equation~\cite{landi2022a} can be readily written as}
\begin{align}
\dot{\f{\rho}} &= -\int_0^\infty d\tau \sum_\nu C_\nu(+\tau) \left[\f{\hat{S}^\nu}(t), \f{\hat{S}^\nu}(t-\tau)\f{\rho}(t)\right]\nn
&\qquad-\int_0^\infty d\tau \sum_\nu C_\nu(-\tau)\left[\f{\rho}(t)\f{\hat{S}^\nu}(t-\tau), \f{\hat{S}^\nu}(t)\right]\,,
\end{align}
where we have denoted the interaction picture by bold symbols and introduced the reservoir correlation function $C_\nu(\tau)=\traceB{e^{+\ii \hat{H}_B^\nu \tau} \hat{B}^\nu e^{-\ii \hat{H}_B^\nu \tau} \hat{B}^\nu \bar\rho_B^\nu}$.
In particular for the non-interacting system Hamiltonian~\eqref{EQ:hamsys} and coupling operators~\eqref{EQ:coupling}, the transformation into the interaction picture can be made very explicit 
\begin{align}\label{EQ:intpic}
  e^{+\ii \hat{H}_S t} \hat{S}^c_\pm e^{-\ii \hat{H}_S t} &= e^{\pm\ii \delta t} \hat{S}^c_\pm\,,\nn
  e^{+\ii \hat{H}_S t} \hat{S}^h_\pm e^{-\ii \hat{H}_S t} &= e^{\pm\ii \Delta t} \hat{S}^h_\pm\,,\nn
  e^{+\ii \hat{H}_S t} \hat{S}^w_\pm e^{-\ii \hat{H}_S t} &= e^{\pm\ii (\Delta-\delta) t} \hat{S}^w_\pm\,.
\end{align}
Thus, writing the Redfield equation in the Schr\"odinger picture, we get (compare e.g.~\cite{liu2021a})
\begin{align}\label{EQ:redfield}
\dot{\rho} &= -\ii \left[\hat{H}_S, \rho(t)\right]-\sum_\nu \int_0^\infty d\tau C_\nu(+\tau) \left[\hat{S}^\nu, \f{\hat{S}^\nu}(-\tau)\rho(t)\right]\nn
&\qquad-\sum_\nu\int_0^\infty d\tau C_\nu(-\tau)\left[\rho(t)\f{\hat{S}^\nu}(-\tau), \hat{S}^\nu\right]\nn
&= -\ii \left[\hat{H}_S, \rho(t)\right]-\sum_\nu\int_0^\infty C_\nu(+\tau) e^{-\ii\Omega_\nu\tau} d\tau \left[\hat{S}^\nu, \hat{S}^\nu_+ \rho\right]\nn
&\qquad-\sum_\nu\int_0^\infty C_\nu(+\tau) e^{+\ii\Omega_\nu\tau} d\tau \left[\hat{S}^\nu, \hat{S}^\nu_-\rho\right]\nn
&\qquad-\sum_\nu\int_0^\infty C_\nu(-\tau) e^{-\ii\Omega_\nu\tau} d\tau \left[\rho \hat{S}^\nu_+, \hat{S}^\nu\right]\nn
&\qquad-\sum_\nu\int_0^\infty C_\nu(-\tau) e^{+\ii\Omega_\nu\tau} d\tau \left[\rho \hat{S}^\nu_-, \hat{S}^\nu\right]\,.
\end{align}
To simplify this expression, one may express the integral prefactors by inserting the Fourier transform of the correlation function $C_\nu(\tau) = \frac{1}{2\pi} \int \gamma_\nu(\omega) e^{-\ii\omega\tau} d\omega$ and afterwards invoking the Sokhotskij-Plemelj theorem 
\begin{align}
\frac{1}{2\pi} \int_0^\infty e^{+\ii\omega\tau} d\tau = \frac{1}{2}\delta(\omega) + \frac{\ii}{2\pi} {\cal P} \frac{1}{\omega}\,,
\end{align}
where ${\cal P}$ denotes the Cauchy principal value.
This allows to write the integrals in terms of Hermitian (real) and anti-Hermitian (imaginary) parts, e.g.
\begin{align}
\int_0^\infty C_\nu(+\tau) e^{+\ii\Omega_\nu\tau} d\tau &= \frac{1}{2} \gamma_\nu(+\Omega_\nu) + \frac{1}{2} \sigma_\nu(+\Omega_\nu)\,,\nn
\int_0^\infty C_\nu(-\tau) e^{+\ii\Omega_\nu\tau} d\tau &= \frac{1}{2} \gamma_\nu(-\Omega_\nu) - \frac{1}{2} \sigma_\nu(-\Omega_\nu)\,,
\end{align}
and analogously for the terms with $\Omega_\nu\to-\Omega_\nu$.
In the equations above, the functions on the r.h.s. are then the even and odd Fourier transforms of the reservoir correlation functions
\begin{align}
\gamma_\nu(\omega) &= \int C_\nu(\tau) e^{+\ii\omega\tau} d\tau\,,\nn
\sigma_\nu(\omega) &= \int C_\nu(\tau) {\rm sgn}(\tau) e^{+\ii\omega\tau} d\tau\,.
\end{align}
The Lamb-shift parts $\sigma_\nu(\pm\Omega_\nu)$ are negligible in comparison to $\hat{H}_S$, such that it is common practice to neglect them.
Eventually, this transforms Eq.~\eqref{EQ:redfield} into Eq.~\eqref{EQ:redfieldmain} in the main text.
The Redfield equation conserves trace and hermiticity, but not necessarily the positivity of the density matrix.
For general operators $\hat{O}$ one can rewrite it in terms of expectation values
\begin{align}
\frac{d}{dt}\expval{\hat{O}} &= +\ii \expval{\left[\hat{H}_S, \hat{O}\right]}\nn
&\qquad+\sum_\nu\frac{\gamma_\nu(-\Omega_\nu)}{2}\Big[\expval{\left[\hat{S}^\nu, \hat{O}\right] \hat{S}_+^\nu}\nn
&\qquad\qquad+\expval{\hat{S}^\nu_-\left[\hat{O}, \hat{S}^\nu\right]}\Big]\nn
&\qquad+\sum_\nu\frac{\gamma_\nu(+\Omega_\nu)}{2}\Big[\expval{\left[\hat{S}^\nu, \hat{O}\right] \hat{S}_-^\nu}\nn
&\qquad\qquad+\expval{\hat{S}^\nu_+\left[\hat{O}, \hat{S}^\nu\right]}\Big]\,.
\end{align}
Thus, it is easy to see that the above equation conserves all operators that commute with both $\hat{H}_S$ and $\hat{S}^\nu$, i.e., if $[\hat{O}, \hat{H}_S]=[\hat{O}, \hat{S}^\nu]=0$, one also has
$\trace{\hat{O} \dot\rho}=0$.
This is the case in the collective limit, where $\hat{S}^\nu\to \hat{J}^\nu$ e.g. for the Casimir operators $\hat{O}\to\hat{C}_i$.

\new{Apart from the Lamb-shift terms, the spectral coupling densities of the reservoirs are always evaluated at their transition frequencies $\delta$, $\Delta$, and $\Delta-\delta$, respectively. 
Thus, for our calculations in the main text (where the Lamb-shift is neglected), the actual form of the spectral coupling density is not relevant.
To confirm that the Lamb-shift contributions are indeed negligible, we employed a special model for the spectral coupling density. 
Using
}
\begin{align}
\Gamma_\nu(\omega) &= \Gamma_\nu \frac{\omega}{\epsilon_\nu} \frac{\delta_\nu^4}{(\omega^2-\epsilon_\nu^2)^2+\delta_\nu^4}\,,
\end{align}
the correlation function $C_\nu(\tau)$ as well as the Lamb-shift contribution 
\begin{align}
\sigma_\nu(\omega) = \frac{\ii}{\pi} {\cal P} \int \frac{\gamma_\nu(\omega')}{\omega-\omega'} d\omega'
\end{align}
may be computed analytically (but yielding extremely lengthy expressions).
\new{To compare with our calculations in the main text, we used resonant reservoirs with $\epsilon_c=\delta$, $\epsilon_h=\Delta$ and $\epsilon_w = \Delta-\delta$
and computed the Redfield current in Fig.~\ref{FIG:currents} with Lamb-shift corrections included (setting $\delta_\nu=\epsilon_\nu$).
For the other parameters chosen as} in Fig.~\ref{FIG:currents} we found the effects of the Lamb-shift contributions to be negligible.

\begin{figure}
    \centering
    \centerline{\includegraphics[width=0.5\textwidth,clip=true]{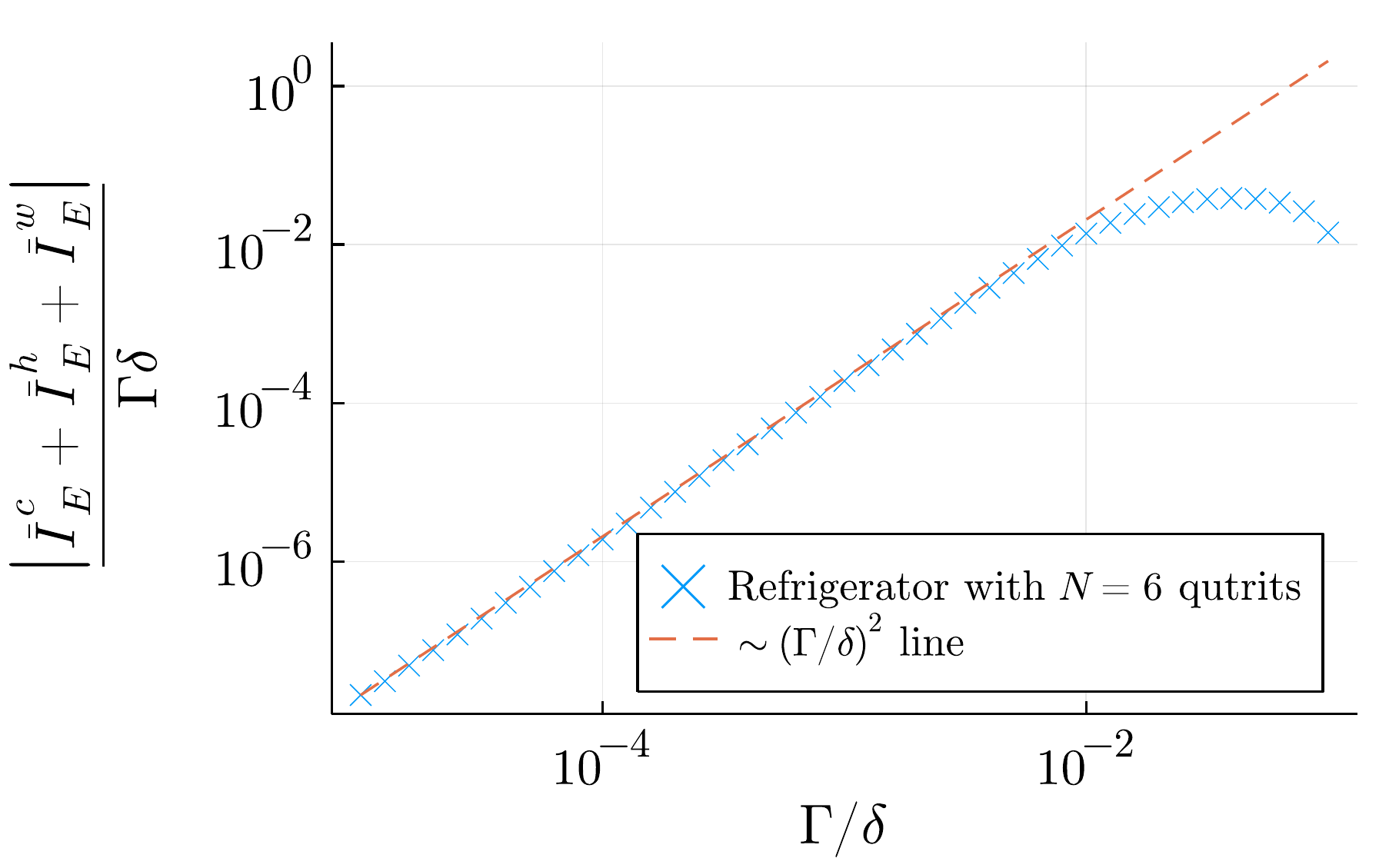}}
    \caption{\new{Double-logarithmic plot of the sum of all stationary Redfield energy currents leaving the reservoirs versus the coupling strength $\Gamma$ for $N=6$ qutrits. The apparent violation of the first law scales as $\Gamma^2$ (dashed line), which is beyond the accuracy of the Redfield equation.    
    Parameters: $\Gamma=\Gamma_c=\Gamma_h=\Gamma_w$, $\Delta=10\delta$, $n_c=10$, $n_h=1$ and $n_w=100$.}}
    \label{FIG:firstlawviolation}
\end{figure}

Similar to the rate equation discussion from App.~\ref{APP:qar1}, the cold reservoir energy current may now be computed using different options:
Taking the view of the system perspective, one may obtain it by computing the system energy balance from~\eqref{EQ:redfieldmain}.
The energy current entering from the cold reservoir then becomes 
\begin{align}
    I_{E,S}^c &= \frac{\Gamma_c n_c}{2} \trace{\hat{H}_S\left(\left[\hat{S}^c_+ \rho, \hat{S}^c\right]+\left[\hat{S}^c, \rho \hat{S}^c_-\right]\right)}\\
    &\quad+\frac{\Gamma_c (1+n_c)}{2} \trace{\hat{H}_S\left(\left[\hat{S}^c, \rho \hat{S}^c_+\right]+\left[\hat{S}^c_- \rho, \hat{S}^c\right]\right)}\,,\nonumber
\end{align}
and analogously for the other reservoirs.
Alternatively, we may microscopically derive energy counting fields~\cite{esposito2009a,liu2021a,landi2022a}, which takes the perspective of energy leaving the reservoir
and yields a tilted Liouvillian ${\cal L}\to {\cal L}(\chi)$.
In Eq.~\eqref{EQ:redfieldmain} in the main text, this would effectively lead to the replacements
$\left(\left[\hat{S}^c, \rho \hat{S}^c_+\right]+\left[\hat{S}^c_-\rho,\hat{S}^c\right]\right)\to \left(\hat{S}^c \rho \hat{S}^c_+ e^{-\ii\delta\chi} + \hat{S}^c_-\rho \hat{S}^c e^{-\ii\delta\chi} - \rho \hat{S}^c_+ \hat{S}^c - \hat{S}^c \hat{S}^c_- \rho\right)$
and the analogous one 
$\left(\left[\hat{S}^c_+ \rho, \hat{S}^c\right] + \left[\hat{S}^c, \rho \hat{S}^c_-\right]\right) \to \left(\hat{S}^c_+ \rho \hat{S}^c e^{+\ii\delta\chi} + \hat{S}^c \rho \hat{S}^c_- e^{+\ii\delta\chi} - \hat{S}^c \hat{S}^c_+ \rho - \rho \hat{S}^c_- \hat{S}^c\right)$.
The current (and noise) can then be computed by the methods in App.~\ref{APP:fcs}.
This yields for the current leaving the cold reservoir
\begin{align}
    I_{E}^c &= -\frac{\Gamma_c\delta(1+n_c)}{2} \trace{\hat{S}^c \rho \hat{S}^c_+ + \hat{S}^c_- \rho \hat{S}^c}\nn
    &\qquad+ \frac{\Gamma_c \delta n_c}{2} \trace{\hat{S}^c_+ \rho \hat{S}^c + \hat{S}^c \rho \hat{S}^c_-}\,.
\end{align}
In general, we have $I_{E}^\nu \neq I_{E,S}^\nu$ for the Redfield equation.
Moreover, while by construction the stationary currents for the system \new{add to zero} 
$\sum_\nu \bar I_{E,S}^\nu = 0$, this is only approximately true in the Redfield approach for the energy currents leaving the reservoir.
\new{In Fig.~\ref{FIG:firstlawviolation} we plot the sum of energy currents leaving the reservoirs (i.e., the deviation from the first law at steady state) versus the coupling strengths to the reservoirs (assumed equal).
One can see that the deviation from energy conservation becomes smaller when the coupling strength is reduced.
Moreover, it scales with $\Gamma^2$ (the Redfield approach is accurate to order $\Gamma$), such this apparent violation can be considered a truncation artifact 
and does not indicate a violation of the first law in the weak-coupling regime, where for our model Redfield and LGKS solutions agree. 
}

\subsection{Secular master equations}\label{APP:lindblad}

Under a secular approximation, Eq.~\eqref{EQ:redfield} simplifies into
\begin{align}
\dot{\f{\rho}} &\approx -\sum_\nu \int_0^\infty d\tau C_\nu(+\tau) e^{-\ii\Omega_\nu\tau} \left[\hat{S}^\nu_-, \hat{S}^\nu_+\f{\rho}(t)\right]\nn
&\qquad-\sum_\nu \int_0^\infty d\tau C_\nu(+\tau) e^{+\ii\Omega_\nu\tau} \left[\hat{S}^\nu_+, \hat{S}^\nu_-\f{\rho}(t)\right]\nn
&\qquad-\sum_\nu \int_0^\infty d\tau C_\nu(-\tau) e^{+\ii\Omega_\nu\tau} \left[\f{\rho}(t)\hat{S}^\nu_-, \hat{S}^\nu_+\right]\nn
&\qquad-\sum_\nu \int_0^\infty d\tau C_\nu(-\tau) e^{-\ii\Omega_\nu\tau} \left[\f{\rho}(t)\hat{S}^\nu_+, \hat{S}^\nu_-\right]\nn
&= -\ii \sum_\nu \left[\frac{\sigma_\nu(-\Omega_\nu)}{2\ii}  \hat{S}^\nu_- \hat{S}^\nu_+
+ \frac{\sigma_\nu(+\Omega_\nu)}{2\ii} \hat{S}^\nu_+ \hat{S}^\nu_-, \f{\rho}\right]\nn
&\qquad+\sum_\nu \gamma_\nu(+\Omega_\nu)\left[\hat{S}^\nu_- \f{\rho} \hat{S}^\nu_+ - \frac{1}{2}\left\{\hat{S}^\nu_+ \hat{S}^\nu_-, \f{\rho}\right\}\right]\nn
&\qquad+\sum_\nu \gamma_\nu(-\Omega_\nu)\left[\hat{S}^\nu_+ \f{\rho} \hat{S}^\nu_- - \frac{1}{2}\left\{\hat{S}^\nu_- \hat{S}^\nu_+, \f{\rho}\right\}\right]\,.
\end{align}
Back in the Schr\"odinger picture, we thus have
\begin{align}\label{EQ:lindbladwithls}
\dot{\rho} &= -\ii \left[\hat{H}_S 
+ \sum_\nu \frac{\sigma_\nu(-\Omega_\nu)}{2\ii} \hat{S}^\nu_- \hat{S}^\nu_+ + \frac{\sigma_\nu(+\Omega_\nu)}{2\ii} \hat{S}^\nu_+ \hat{S}^\nu_-, \f{\rho}\right]\nn
&\qquad+\sum_\nu \gamma_\nu(+\Omega_\nu)\left[\hat{S}^\nu_- \rho \hat{S}^\nu_+ - \frac{1}{2}\left\{\hat{S}^\nu_+ \hat{S}^\nu_-, \rho\right\}\right]\nn
&\qquad+\sum_\nu \gamma_\nu(-\Omega_\nu)\left[\hat{S}^\nu_+ \rho \hat{S}^\nu_- - \frac{1}{2}\left\{\hat{S}^\nu_- \hat{S}^\nu_+, \rho\right\}\right]\,,
\end{align}
which upon neglecting the Lamb shift $\sigma_\nu(\pm\Omega_\nu)$ reduces to Eq.~\eqref{EQ:lindblad} from the main text.
This equation is of LGKS form and thermalizes the system when all reservoirs are at the same equilibrium temperature.
Furthermore, when evaluating only the populations of $\rho$ in the system energy eigenbasis for situations where the system Hamiltonian is non-degenerate (at least within a conserved subspace), we obtain the simple Pauli-type rate equation as exemplified in App.~\ref{APP:pauli}.

In the collective limit, the associated effective non-Hermitian Hamiltonian
\begin{align}
    H_{\rm eff} = \hat{H}_S - \frac{\ii}{2} \sum_\nu \left[\Gamma_\nu (1+n_\nu) \hat{J}^\nu_+ \hat{J}^\nu_- + \Gamma_\nu n_\nu \hat{J}^\nu_+ \hat{J}^\nu_-\right]
\end{align}
is diagonal in the maximum symmetry subspace $H_{\rm eff} \ket{M;m} = \lambda_{Mm} \ket{M;m}$.
Its eigenvalues can be calculated analytically via~\eqref{EQ:ladder_coefficients}, and their imaginary part -- related to the waiting time distribution~\cite{dalibard1989a} between any two jump events~\cite{schaller2022a}  -- indicates that for fixed thermal reservoirs $n_\nu$ and coupling strengths $\Gamma_\nu$, the waiting time is minimized for $M\approx m\approx N/3$ (maximum engine activity).
Specifically, from the imaginary part of
\begin{align}
    \lambda_{\frac{N}{3},\frac{N}{3}} = \frac{N}{3}(\Delta+\delta) - \frac{\ii}{2} \frac{N}{3}\left(\frac{N}{3}+1\right) \left[\sum_\nu \Gamma_\nu (1+2n_\nu)\right]
\end{align}
we can conclude that the lifetime of the states participating in the central cycle scales as $\tau \propto N^{-2}$.

The currents can now be computed in analogy to the previous section, i.e., either by the energy flowing into the system from the cold reservoir
\begin{align}
    I_{E,S}^c &= \Gamma_c(1+n_c) \trace{\hat{H}_S \left[\hat{S}^c_- \rho \hat{S}^c_+ - \frac{1}{2}\left\{\hat{S}^c_+ \hat{S}^c_-, \rho\right\}\right]}\nn
&\qquad+\Gamma_c n_c \trace{\hat{H}_S \left[\hat{S}^c_+ \rho \hat{S}^c_- - \frac{1}{2}\left\{\hat{S}^c_- \hat{S}^c_+, \rho\right\}\right]}\,,
\end{align}
or by assessing the energy flow out of the cold reservoir with counting fields $\hat{S}^c_- \rho \hat{S}^c_+\to \hat{S}^c_- \rho \hat{S}^c_+ e^{-\ii\delta\chi}$
and $\hat{S}^c_+ \rho \hat{S}^c_-\to \hat{S}^c_+ \rho \hat{S}^c_- e^{+\ii\delta\chi}$ and then using the methods of Sec.~\ref{APP:fcs}.
The current leaving the cold reservoir can then be obtained via $I_{E}^c(t) = -\ii \trace{{\cal L}'(0)\vec\rho(t)}$ and becomes
\begin{align}
    I_{E}^c &=  \Gamma_c \delta n_c \trace{\hat{S}^c_+ \rho \hat{S}^c_-} - \Gamma_c\delta (1+n_c)\trace{\hat{S}^c_- \rho \hat{S}^c_+}\,.
\end{align}
In contrast to the Redfield case discussed above, we find $I_{E}^\nu=I_{E,S}^\nu$, such that we can unambiguously write the first law of thermodynamics as 
\begin{align}
    \frac{d}{dt} \trace{\hat{H}_S \rho} = \sum_\nu I_E^\nu(t)\,.
\end{align}
With using Spohn's inequality~\cite{spohn1978b}, one can also establish the second law of thermodynamics
\begin{align}
    \frac{d}{dt} \trace{-\rho \ln \rho} -\sum_\nu \beta_\nu I_E^\nu(t) \ge 0\,.
\end{align}
These bound the coefficient of performance of the QAR by its Carnot value.

\subsection{Pauli rate equation}\label{APP:pauli}

For the fully symmetric case, the derivation of a rate equation becomes particularly simple: 
We can formally insert a diagonal density matrix in the fully symmetric subspace
\begin{align}
\rho = \sum_{M,m} P_{Mm} \ket{M;m}\bra{M;m}\,,
\end{align}
into~\eqref{EQ:lindblad}, which with the help of Eq.~\eqref{EQ:ladder_coefficients} yields
\begin{align}
\dot P_{Mm} &= \gamma_c(-\delta)(N-M-m+1)m P_{M,m-1}\nn
&\qquad-\gamma_c(-\delta) (N-M-m)(m+1) P_{M,m}\nn
&\qquad+\gamma_c(+\delta) (N-M-m)(m+1) P_{M,m+1}\nn
&\qquad-\gamma_c(+\delta) (N-M-m+1)m P_{M,m}\nn
&\qquad+\gamma_h(-\Delta) (N-M-m+1)M P_{M-1,m}\nn
&\qquad-\gamma_h(-\Delta) (N-M-m)(M+1) P_{M,m}\nn
&\qquad+\gamma_h(+\Delta) (N-M-m)(M+1) P_{M+1,m}\nn
&\qquad-\gamma_h(+\Delta) (N-M-m+1)M P_{M,m}\nn
&\qquad+\gamma_w(-\Delta+\delta) M(m+1) P_{M-1,m+1}\nn
&\qquad-\gamma_w(-\Delta+\delta) (M+1)m P_{M,m}\nn
&\qquad+\gamma_w(+\Delta-\delta) (M+1)m P_{M+1,m-1}\nn
&\qquad-\gamma_w(+\Delta-\delta) M(m+1) P_{M,m}\,.
\end{align}
This is precisely the Pauli-type rate equation~\eqref{EQ:paulirate} in the main text, 
where we note that rates outside allowed states do naturally vanish.
In particular, we can identify the transitions that increase the system energy as 
\begin{align}\label{EQ:rateup}
R_{(M,m+1),(M,m)} &= \Gamma_c n_c \cdot (N-M-m)(m+1)\,,\nn
R_{(M+1,m),(M,m)} &= \Gamma_h n_h \cdot (N-M-m)(M+1)\,,\nn
R_{(M+1,m-1),(M,m)} &= \Gamma_w n_w \cdot m (M+1)\,,
\end{align}
while transitions that decrease it are always a bit larger due to local detailed balance
\begin{align}\label{EQ:ratedn}
R_{(M,m),(M,m+1)} &=\Gamma_c (1+n_c) \cdot (N-M-m)(m+1)\,,\nn
R_{(M,m),(M+1,m)} &= \Gamma_h (1+n_h) \cdot (N-M-m)(M+1)\,,\nn
R_{(M,m),(M+1,m-1)}  &= \Gamma_w (1+n_w) \cdot m (M+1)\,.
\end{align}

Currents can be obtained in analogy to the case $N=1$ discussed in App.~\ref{APP:qar1}:
Either one computes them via the energy balance of the system which, for a system subject to the rate equation $\dot{P}_i = \sum_{ij} \sum_\nu \left[R_{ij}^\nu P_j - R_{ji}^\nu P_i\right]$ with $R_{ij}^\nu$ denoting the transition rate between system energy eigenstate $j\to i$ triggered by reservoir $\nu$, generically yields the expression
$I_E^\nu = \sum_{ij} (E_i-E_j) R_{ij}^\nu P_j$. 
Alternatively, one may also insert counting fields (or use the ones from the LGKS description), which effectively corresponds to the replacement
$\Gamma_c n_c \to \Gamma_c n_c e^{+\ii\delta\chi}$ and $\Gamma_c (1+n_c) \to \Gamma_c (1+n_c) e^{-\ii\delta\chi}$ in the off-diagonal matrix elements of the rate matrix and then use the methods of App.~\ref{APP:fcs}.
As with the previous section, these two approaches are equivalent and thermodynamically consistent.

For multi-qutrit systems the exact computation of the steady state is cumbersome already for the collective limit and a Pauli rate equation~\eqref{EQ:paulirate}, such that we do not provide analytic results for the currents.
However, in App.~\ref{APP:tightcoupling} we perform a cycle decomposition to establish tight-coupling relations between the currents.

\subsection{Coarse-grained rate equation}\label{APP:coarsegraining}

For the case where the work reservoir is infinitely hot $n_w\to\infty$, the associated green transitions in Fig.~\ref{FIG:threels} become dominant, 
and along them the populations become approximately equal (but remain still conditioned on the total number of excitations).
This allows us to coarse-grain~\cite{pigolotti2008a,esposito2012b} the populations by summarizing all states with same total number of excitations
\begin{align}
Q_n = \sum_{M,m} \delta_{M+m,n} P_{M,m}\,.
\end{align}
As the states connected by work reservoir transitions equilibrate much faster in this limit, we have for their conditional probability
\begin{align}
\lim_{n_w\to\infty} \left.\frac{P_{M,m}}{Q_n}\right|_{M+m=n} = \frac{1}{n+1}\,.
\end{align}
Summing over the respective states of Eq.~\eqref{EQ:paulirate} allows us to obtain a coarse-grained one-dimensional Markovian rate equation
\begin{align}
\dot Q_n &= {\sum_{M,m}\delta_{M+m,n}}\Big[ 
\gamma_c(-\delta)(N-M-m+1)m \frac{P_{M,m-1}}{Q_{n-1}} Q_{n-1}\nn
&\qquad-\gamma_c(-\delta) (N-M-m)(m+1) \frac{P_{M,m}}{Q_n} Q_n\nn
&\qquad+\gamma_c(+\delta) (N-M-m)(m+1) \frac{P_{M,m+1}}{Q_{n+1}} Q_{n+1}\nn
&\qquad-\gamma_c(+\delta) (N-M-m+1)m \frac{P_{M,m}}{Q_n} Q_n\nn
&\qquad+\gamma_h(-\Delta) (N-M-m+1)M \frac{P_{M-1,m}}{Q_{n-1}} Q_{n-1}\nn
&\qquad-\gamma_h(-\Delta) (N-M-m)(M+1) \frac{P_{M,m}}{Q_n} Q_n\nn
&\qquad+\gamma_h(+\Delta) (N-M-m)(M+1) \frac{P_{M+1,m}}{Q_{n+1}} Q_{n+1}\nn
&\qquad-\gamma_h(+\Delta) (N-M-m+1)M \frac{P_{M,m}}{Q_n} Q_n\nn
&\qquad+\gamma_w(-\Delta+\delta) M(m+1) \frac{P_{M-1,m+1}}{Q_n} Q_n\nn
&\qquad-\gamma_w(-\Delta+\delta) (M+1)m \frac{P_{M,m}}{Q_n} Q_n\nn
&\qquad+\gamma_w(+\Delta-\delta) (M+1)m \frac{P_{M+1,m-1}}{Q_n} Q_n\nn
&\qquad-\gamma_w(+\Delta-\delta) M(m+1) \frac{P_{M,m}}{Q_n} Q_n\Big]\nn
&\stackrel{n_w\to\infty}{\approx}
R_{n,n+1}^{\rm cg} Q_{n+1} + R_{n,n-1}^{\rm cg} Q_{n-1}\nn
&\qquad - [R_{n-1,n}^{\rm cg} + R_{n+1,n}^{\rm cg}] Q_n\,,
\end{align}
where $n\in\{0,1,\ldots,N\}$ denotes the total number of excitations and we have inserted the conditional probabilities.
In the coarse-grained rates, the mesostate-internal transitions due to the work reservoir cancel out, whereas the other reservoirs determine the
transitions between mesostates
\begin{align}
R_{n,n+1}^{\rm cg} &= \sum_{M,m} \frac{\delta_{M+m,n}}{n+2} \Big[\Gamma_c (1+n_c) (N-M-m)(m+1)\nn
&\qquad+\Gamma_h (1+n_h) (N-M-m)(M+1)\Big]\,,\nn
R_{n,n-1}^{\rm cg} &= \sum_{M,m} \frac{\delta_{M+m,n}}{n} \Big[\Gamma_c n_c (N-M-m+1)m\nn
&\qquad+\Gamma_h n_h (N-M-m+1)M\Big]\,,
\end{align}
which evaluate to Eq.~\eqref{EQ:cgrate_rates} in the main text.
Note that the case $R_{0,-1}^{\rm cg}=0$ has to be treated separately as the state with zero excitations does not need to be coarse-grained.
%
%
%
As a sanity check, we remark that for $N=1$, the proper coarse-grained rates (which one gets by applying an analogous procedure to the rate matrices in Sec.~\ref{APP:qar1}) for a single QAR are reproduced.  
We note that the resulting effective rates for cold and hot reservoir no longer obey local detailed balance, which allows energy to flow out of the cold reservoir.
The simple tri-diagonal form of this effective rate equation has the advantage that the stationary state can be computed analytically, since it obeys
\begin{align}
\bar Q_n = \frac{R_{n,n-1}^{\rm cg}}{R_{n-1,n}^{\rm cg}} \frac{R_{n-1,n-2}^{\rm cg}}{R_{n-2,n-1}^{\rm cg}} \cdot \ldots \cdot \frac{R_{1,0}^{\rm cg}}{R_{0,1}^{\rm cg}} \bar Q_0\,,\qquad
\sum_n \bar Q_n = 1\,.
\end{align}
Making the ratios explicit with~\eqref{EQ:cgrate_rates}, we can determine $\bar Q_0$ from the normalization condition and from that write the stationary cooling current with $\bar n = \frac{\Gamma_c n_c + \Gamma_h n_h}{\Gamma_c+\Gamma_h}$ and $\alpha_{\bar n} = \frac{\bar n}{\bar n +1}$ and $R_{ij}^c \equiv R_{ij}^{\rm cg}|_{\Gamma_h\to 0}$
\begin{align}\label{EQ:cg_analytical_current}
    \bar I_E^c &= \delta \sum_{n=1}^{N} R_{n,n-1}^c \bar Q_{n-1} - \delta \sum_{n=0}^{N-1} R_{n,n+1}^c \bar Q_{n+1}\nn
&= \frac{\Gamma_c \delta}{2} \sum_{n=0}^N \Big[(n+2) (N-n) n_c\nn
    &\qquad\qquad- n (N-n+1)(1+n_c)\Big] \bar Q_n\nn
    &= \frac{\Gamma_c \Gamma_h \delta}{2(\Gamma_c+\Gamma_h)} (n_c-n_h) f_N(\bar n)\,,\nn
    f_N(\bar n) &\equiv \frac{g_N(\bar n)+\alpha_{\bar n}^{N+1} h_N(\bar n)}{\bar n + 1 - \alpha_{\bar n}^{N+1} \left[2+\bar n + N\right]}\,,\nn
    g_N(\bar n) &\equiv 2(N-3\bar n)(\bar n + 1)\,,\nn
    h_N(\bar n) &\equiv N^2+(5+4\bar n)N+6(\bar n + 1)^2
    \,.
\end{align}
In the current, only the term $(n_c-n_h)$ can turn negative, such that we recover the original cooling condition for $n_w\to\infty$ given in~\eqref{EQ:coolingcond0}.
Apart from that, the scaling factor $f_N(\bar n)>0$ can be analyzed for various limits:
First of all, we have $f_1(\bar n) = \frac{2}{1+3 \bar n}$, which yields the same current as~\eqref{EQ:current_qar1}.
Second, for very large $N\gg \bar n$, we can drop the terms with powers of $\alpha_{\bar n} < 1$, such that $f_N(\bar n) \to 2 N$, and the current eventually will scale just linearly $\bar I_E^c \approx \frac{\Gamma_c \Gamma_h \delta}{\Gamma_c+\Gamma_h} (n_c-n_h)N$.
Third, for $N \ll \bar n$ we have $f_N(\bar n) \to \frac{N(N+3)}{6\bar n}$, and we maintain a quadratic scaling for the current
$\bar I_E^c \approx \frac{\Gamma_c \Gamma_h \delta (n_c-n_h)}{12(\Gamma_c n_c + \Gamma_h n_h)} N(N+3)$.
The crossover system size $N^*$ between these regimes can be found by simply equating the limits and yields $N^* = 12 \bar n - 3$.
These limits can be seen well in Fig.~\ref{FIG:currents} (dashed \new{magenta} curves).
The crossover between two scaling regimes is thus quite analogous to previous results for collective qubit systems~\cite{vogl2011a}.
We can see that the current becomes maximal when $n_h\to 0$ (which can be reached by $\Delta\to\infty$ and then implies $n_c \to \frac{\Gamma_c+\Gamma_h}{\Gamma_c} \bar n$).
Then, we can numerically maximize $(n_c-n_h) f_N(\bar n) \to \frac{\Gamma_c+\Gamma_h}{\Gamma_c} \bar{n} f_N(\bar n)$ as a function of $\bar n$ only.
The position of this maximum is for large $N$ roughly at $\bar n \approx N/6$, and the current scales quadratically in $N$ at this maximum.

\subsection{Interacting Pauli rate equation}\label{APP:pauli_int}

Pauli rate equations can also be derived for interactions present in the system Hamiltonian $\hat{H}_S$.
If these interactions can be expressed by the permutationally symmetric operators $\hat{C}_2$, $\hat{N}_\delta$ and $\hat{N}_\Delta$ (or equivalently by $\sum_\alpha \hat{J}_\alpha^2$, $\hat{J}_3$ and $\hat{J}_8$) like in Eq.~\eqref{EQ:ham_penalty}, it follows that the same eigenstates can be used for the representation of the problem, and only the eigenvalues change.
Labelling the eigenstates of $\hat{H}_S$ with the multi-index $i$, the Pauli rate equation will then generically have the structure~\cite{breuer2002}
\begin{align}
    \dot{P}_i &= \sum_\nu \sum_j \left[R_{ij}^\nu P_j - R_{ji}^\nu P_i\right]\,,\nn
    R_{ij}^\nu &= \gamma_\nu(E_j-E_i) \abs{\bra{i} \hat{S}^\nu \ket{j}}^2\,.
\end{align}
The inherent local detailed balance for the transition rates then exponentially suppresses excitations into undesired Casimir subspaces and undesired excitation numbers, whereas relaxation from the excited eigenvalues down to the desired most productive cycle is still possible.
Thus, for $N=3k+1$ with integer $k$ and sufficiently large penalty parameters $\beta_\nu \alpha_{C/P}\gg 1$, we can neglect the excited states and constrain our considerations to the states from the maximum symmetry (Casimir) sector with $\expval{\hat{C}_2} = N(N+3)/3$:
$\ket{k;k}$, $\ket{k;k+1}$,  and $\ket{k+1;k}$.
These three states then have the energies 
\begin{align}
    E_0 &= k(\Delta+\delta)+\alpha_P/3\,,\nn
    E_1 &= k\Delta+(k+1)\delta+\alpha_P/3\,,\nn
    E_2 &= (k+1)\Delta + k\delta+\alpha_P/3\,,
\end{align}
such that the penalty $\alpha_P$ cancels out in their differences.
As the eigenstates $\ket{M;m}$ remain the same, the quadratic enhancement from Eq.~\eqref{EQ:ladder_coefficients} is preserved, and we obtain the current~\eqref{EQ:current_qar1} with $\Gamma_\nu \to \Gamma_\nu \left(\frac{N+2}{3}\right)^2$, leading to the corresponding enhancement~\eqref{EQ:current_boost} in the main text.


\section{Bosonization}\label{APP:bosonization}

The $8$ operators $\{\hat{J}_3, \hat{J}_8, \hat{J}^h_+, \hat{J}^h_-, \hat{J}^c_+, \hat{J}^c_-, \hat{J}^w_+, \hat{J}^w_-\}$ inherit a closed algebra from the associated single-qutrit versions
\begin{align}
\left[\hat{J}_3, \hat{J}^h_+\right] &= \frac{\hat{J}^h_+}{2}\,,\;
\left[\hat{J}_3, \hat{J}^c_+\right] = -\frac{\hat{J}^c_+}{2}\,,\;
\left[\hat{J}_3, \hat{J}^w_+\right] = \hat{J}^w_+\,,\nn
\left[\hat{J}_8, \hat{J}^h_+\right] &= \frac{\sqrt{3} \hat{J}^h_+}{2}\,,\;
\left[\hat{J}_8, \hat{J}^c_+\right] = \frac{\sqrt{3} \hat{J}^c_+}{2}\,,\nn
\left[\hat{J}^h_+, \hat{J}^h_-\right] &= \sqrt{3}\hat{J}_8+\hat{J}_3\,,\;
\left[\hat{J}^c_+, \hat{J}^c_-\right] = \sqrt{3}\hat{J}_8-\hat{J}_3\,,\nn
\left[\hat{J}^w_+, \hat{J}^w_-\right] &= 2 \hat{J}_3\,,\nn
\left[\hat{J}^h_+, \hat{J}^c_-\right] &= \hat{J}^w_+\,,\;
\left[\hat{J}^h_+, \hat{J}^w_-\right] = -\hat{J}^c_+\,,\nn
\left[\hat{J}^c_+, \hat{J}^w_+\right] &= -\hat{J}^h_+\,.
\end{align}
Other independent (which do not follow from Hermitian conjugation) commutators just vanish.

\subsection{Four-mode Holstein-Primakoff transform}

The above commutation relations can be realized with four bosonic modes with annihilation operators $\hat{a}_\Delta$, $\hat{a}_\delta$, $\hat{b}_\Delta$, and $\hat{b}_\sigma$ obeying the usual bosonic commutation relations by using a generalization~\cite{providencia2006a} of the Holstein-Primakoff-transform
\begin{align}\label{EQ:4mode_holstein_primakoff}
\hat{J}_3 &= \hat{b}_\sigma^\dagger \hat{b}_\sigma +\frac{1}{2} \hat{b}_\Delta^\dagger \hat{b}_\Delta + \frac{1}{2}\left(\hat{a}_\Delta^\dagger \hat{a}_\Delta - \hat{a}_\delta^\dagger \hat{a}_\delta\right)-\frac{N_b}{2}\,,\nn
\hat{J}_8 &= \frac{\sqrt{3}}{2} \hat{b}_\Delta^\dagger \hat{b}_\Delta+\frac{\sqrt{3}}{2} \left(\hat{a}_\Delta^\dagger \hat{a}_\Delta + \hat{a}_\delta^\dagger \hat{a}_\delta\right) - \frac{N_a}{\sqrt{3}}-\frac{N_b}{2\sqrt{3}}\,,\nn
\hat{J}^h_+ &= \hat{a}_\Delta^\dagger \sqrt{N_a - \hat{a}_\Delta^\dagger \hat{a}_\Delta - \hat{a}_\delta^\dagger \hat{a}_\delta}\nn
&\qquad+\hat{b}_\Delta^\dagger\sqrt{N_b - \hat{b}_\Delta^\dagger \hat{b}_\Delta - \hat{b}_\sigma^\dagger \hat{b}_\sigma}\,,\nn
\hat{J}^c_+ &= \hat{a}_\delta^\dagger \sqrt{N_a - \hat{a}_\Delta^\dagger \hat{a}_\Delta - \hat{a}_\delta^\dagger \hat{a}_\delta}+\hat{b}_\Delta^\dagger \hat{b}_\sigma\,,\nn
\hat{J}^w_+ &= \hat{a}_\Delta^\dagger \hat{a}_\delta-\hat{b}_\sigma^\dagger\sqrt{N_b-\hat{b}_\Delta^\dagger \hat{b}_\Delta - \hat{b}_\sigma^\dagger \hat{b}_\sigma}\,.
\end{align}
Here, the integer numbers $N_a\ge 0$ and $N_b\ge 0$ determine the physically admissable states, i.e., they have to be adjusted to match the behaviour of the collective qutrit operators such as e.g. the Hamiltonian
\begin{align}\label{EQ:hamsysboson}
    \hat{H}_S &= \Delta \left[\hat{a}_\Delta^\dagger \hat{a}_\Delta + \hat{b}_\Delta^\dagger \hat{b}_\Delta +\hat{b}_\sigma^\dagger \hat{b}_\sigma + \frac{N-N_a-2N_b}{3}\right]\nn
    &\qquad+\delta \left[\hat{a}_\delta^\dagger \hat{a}_\delta - \hat{b}_\sigma^\dagger \hat{b}_\sigma + \frac{N-N_a+N_b}{3}\right]\,.
\end{align}
In particular, the Fock states $\ket{M,m,Q,q}$ with $M$, $m$, $Q$, and $q$ denoting the eigenvalues of $\hat{a}_\Delta^\dagger \hat{a}_\Delta$, $\hat{a}_\delta^\dagger \hat{a}_\delta$, $\hat{b}_\Delta^\dagger \hat{b}_\Delta$, and $\hat{b}_\sigma^\dagger \hat{b}_\sigma$, respectively, are physically admissable when 
\begin{align}
    &[(N_a = M+m) \lor (Q=0)] \land\nn
    &[(N_b = Q+q) \lor (M=0)] \land\nn
    &[(q=0) \lor (m=0)]\,.
\end{align}
For example, the special case $Q=q=N_b=0$ fulfils these conditions and admits an even simpler representation of the algebra with just two bosonic modes, which we discuss below, and that corresponds to the fully symmetric subspace discussed in the main text.
For the subspace with second largest Casimir operator eigenvalue one has to use the four-boson representation instead.
As one always has $N_a+N_b \le N$ and $0 \le M+m \le N_a$ as well as $0 \le Q+q \le N_b$, it follows that largest Clebsch-Gordan coefficients (and therefore the largest currents) may originate from the fully symmetric subspace.

\subsection{Two-mode Holstein-Primakoff transform}

When $Q=q=N_b=0$, the transformation requires only two bosonic modes~\cite{klein1991a,hayn2011a} with a non-negative integer $N_a$
\begin{align}\label{EQ:holstein_primakoff}
\hat{J}_3 &= \frac{1}{2} \left(\hat{a}_\Delta^\dagger \hat{a}_\Delta - \hat{a}_\delta^\dagger \hat{a}_\delta\right)\,,\nn
\hat{J}_8 &= \frac{\sqrt{3}}{2} \left(\hat{a}_\Delta^\dagger \hat{a}_\Delta + \hat{a}_\delta^\dagger \hat{a}_\delta\right) - \frac{N_a}{\sqrt{3}}\,,\nn
\hat{J}^h_+ &= \hat{a}_\Delta^\dagger \sqrt{N_a - (\hat{a}_\Delta^\dagger \hat{a}_\Delta + \hat{a}_\delta^\dagger \hat{a}_\delta)}\,,\nn
\hat{J}^c_+ &= \hat{a}_\delta^\dagger \sqrt{N_a - (\hat{a}_\Delta^\dagger \hat{a}_\Delta + \hat{a}_\delta^\dagger \hat{a}_\delta)}\,,\nn
\hat{J}^w_+ &= \hat{a}_\Delta^\dagger \hat{a}_\delta\,,
\end{align}
and analogously for the lowering operators.
It turns out that the fully symmetric subspace can be covered by the choice $N_a=N$ (number of qutrits).
Then, the fully symmetric states discussed in the main text are equivalent to the Fock state representation with just two bosonic modes
$\ket{M;m} \equiv \ket{M,m,0,0}$.

The system Hamiltonian~\eqref{EQ:hamsysboson} then simply assumes the form $\hat{H}_S = \Delta \hat{a}_\Delta^\dagger \hat{a}_\Delta + \delta \hat{a}_\delta^\dagger \hat{a}_\delta$.
As a sanity check, representing the Casimir operator in terms of the bosons we get the maximum eigenvalue valid for the fully symmetric subspace
$\expval{\hat{C}_2} = \frac{N(N+3)}{3}$.
From the bosonic properties it is then straightforward to compute the Clebsch-Gordan coefficients~\eqref{EQ:ladder_coefficients} of the symmetric subspace discussed in the main text.

\subsection{Master equation for large $N$}

The master equation~\eqref{EQ:lindblad} can in the symmetric subspace and for large $N$ such that $\expval{a_\nu^\dagger a_\nu} \ll N$ be simplified by expanding the roots in~\eqref{EQ:holstein_primakoff}, which yields
\begin{align}
\dot\rho &= -\ii \left[\Delta \hat{a}_\Delta^\dagger \hat{a}_\Delta + \delta \hat{a}_\delta^\dagger \hat{a}_\delta, \rho\right]\\
&\qquad+N\Gamma_c(1+n_c) \left[\hat{a}_\delta \rho \hat{a}_\delta^\dagger - \frac{1}{2} \left\{\hat{a}_\delta^\dagger \hat{a}_\delta, \rho\right\}\right]\nn
&\qquad+N\Gamma_c n_c \left[\hat{a}_\delta^\dagger \rho \hat{a}_\delta - \frac{1}{2} \left\{\hat{a}_\delta \hat{a}_\delta^\dagger, \rho\right\}\right]\nn
&\qquad+N\Gamma_h(1+n_h) \left[\hat{a}_\Delta \rho \hat{a}_\Delta^\dagger - \frac{1}{2} \left\{\hat{a}_\Delta^\dagger \hat{a}_\Delta, \rho\right\}\right]\nn
&\qquad+N\Gamma_h n_h \left[\hat{a}_\Delta^\dagger \rho \hat{a}_\Delta - \frac{1}{2} \left\{\hat{a}_\Delta \hat{a}_\Delta^\dagger, \rho\right\}\right]\nn
&\qquad+\Gamma_w(1+n_w) \left[\hat{a}_\delta^\dagger \hat{a}_\Delta \rho \hat{a}_\Delta^\dagger \hat{a}_\delta - \frac{1}{2}\left\{\hat{a}_\Delta^\dagger \hat{a}_\delta \hat{a}_\delta^\dagger \hat{a}_\Delta, \rho\right\}\right]\nn
&\qquad+\Gamma_w n_w \left[\hat{a}_\Delta^\dagger \hat{a}_\delta \rho \hat{a}_\delta^\dagger \hat{a}_\Delta - \frac{1}{2}\left\{\hat{a}_\delta^\dagger \hat{a}_\Delta \hat{a}_\Delta^\dagger \hat{a}_\delta, \rho\right\}\right]\,.\nonumber
\end{align}
In absence of the work reservoir $\Gamma_w\to 0$, we find that relaxation to the steady state is therefore scaling with $N t$
\begin{align}
\expval{\hat{a}_\delta^\dagger \hat{a}_\delta}_t = N_\delta^0 e^{-N \Gamma_c t} + n_c (1-e^{-N \Gamma_c t})\,,
\end{align}
and analogously for the hot reservoir. 
The above also manifests superradiant decay as the relaxation time is inversely proportional to $N$.
However, the above master equation is valid only in regimes where $\expval{a_\nu^\dagger a_\nu} \ll N$, such that we cannot describe the regime of boosted cooling power with it.


\section{Full Counting Statistics}\label{APP:fcs}

The starting point for the determination of currents and their fluctuations is a generalized (or tilted) Liouvillian (or rate matrix) equation of the form
\begin{align}
    \dot{\rho} = {\cal L}(\chi) \rho\,,
\end{align}
where $\rho$ is the vectorized part of interest of the density matrix (this could be the complete system density matrix or just the populations of a relevant subspace) and ${\cal L}(\chi)$ the matrix representing the corresponding part of the Redfield or LGKS dissipator or of the rate matrix that depends on the counting field $\chi$.
In the vectorized space, the trace is computed via $\trace{\rho} = \vec{\f{1}}^{\rm T} \cdot \vec{\rho}$, and trace conservation then implies
that $\trace{{\cal L}(0) \sigma} = 0$ for any operator $\sigma$.

The moments of the conjugate variable $n$ to the counting field can be obtained by taking derivatives of the moment-generating function
\begin{align}
    M(\chi,t) = \trace{e^{{\cal L}(\chi) t} \rho_0}
\end{align}
via $\expval{n^\alpha}_t = (-\ii \partial_\chi)^\alpha M(\chi,t)|_{\chi=0}$.
Analogously, one may obtain cumulants from the cumulant-generating function
$C(\chi,t) = \ln M(\chi,t)$ by acting with the corresponding derivatives on it.
In the long-term limit and for systems with a unique stationary state, one can show that $C(\chi,t)\to \lambda(\chi) t$, where $\lambda(\chi)$ is the dominant eigenvalue of ${\cal L}(\chi)$ (the one with largest real part, that fulfils $\lambda(0)=0$).
We are interested in the long-term limit of the lowest two cumulants of the current, i.e., in the current $\bar I = \lim_{t\to\infty}\frac{d}{dt} \expval{n}_t = -\ii \lambda'(0)$ and the noise $\bar S_I=\lim_{t\to\infty}\frac{d}{dt}\left[\expval{n^2}_t-\expval{n}_t^2\right] = -\lambda''(0)$.
Unfortunately, for our problems, the dominant eigenvalue is not analytically known.
Second order perturbation theory for non-Hermitian matrices is rather non-trivial~\cite{buth2004a} and for larger $N$ numerical differentiation~\cite{liu2021a} is not stable.
Therefore, to compute the stationary current and noise, we use a different approach~\cite{landi2022a}, derived from the counting statistics of time-dependent conductors~\cite{benito2016b}.
Trace conservation implies that the stationary current can be obtained via
\begin{align}
    \bar I = -\ii \trace{{\cal L}'(0) \bar\rho}\,,
\end{align}
where $\bar\rho$ is the solution to the equation ${\cal L}(0)\bar\rho=0$ normalized to $\trace{\bar\rho}=1$ (i.e., the steady state).
With that, we can also compute the stationary noise
\begin{align}
    \bar S_{I} = -\trace{{\cal L}''(0)\bar\rho}-2\ii \trace{{\cal L}'(0)\bar\sigma}\,,
\end{align}
where the auxiliary variable $\bar\sigma$ is the solution to the equation
${\cal L}(0)\bar\sigma = \ii {\cal L}'(0)\bar\rho + \bar I \bar \rho$ normalized to $\trace{\bar\sigma}=0$.


\section{Tight-coupling of energy currents}\label{APP:tightcoupling}

The Pauli rate equation~\eqref{EQ:paulirate} has the form
$\dot P_i = \sum_\nu \sum_j \left(R_{ij}^\nu P_j - R_{ji}^\nu P_i\right)$, 
where $P_i$ are occupation probabilities of energy eigenstate $i$ (which in our state corresponds to the state $\ket{M;m}$ of the fully symmetric sector) and $\nu \in \{c,h,w\}$ labels the reservoirs.
We can split the stationary populations into the contributions from the individual cycles that couple to state $i$
\begin{align}
    \bar P_i = \sum_{\cal C} \bar P_i^{\cal C}\,,
\end{align}
where e.g. in Fig.~\ref{FIG:threels} right panel the corner states take part in just one cycle, states on the facets take part in three cycles, and states in the interior take part in six cycles.
Then, the steady-state condition can be rewritten as
\begin{align}
    0 = \sum_{\cal C} \left[\sum_\nu \sum_j \left(R_{ij}^{\nu,\cal C} \bar P_j^{\cal C} - R_{ji}^{\nu,\cal C} \bar P_i^{\cal C}\right)\right]\,,
\end{align}
and fulfilling it for every cycle obviously fulfills the complete steady-state condition.
The individual conditions for a normally oriented ($+$) cycle read
\begin{align}\label{EQ:steadystateplus}
    0 &= \Gamma_h^{\cal C} (1+n_h) \bar P_2^{\cal C} + \Gamma_c^{\cal C} (1+n_c) \bar P_1^{\cal C}\nn
    &\qquad- \left(\Gamma_h^{\cal C} n_h + \Gamma_c^{\cal C} n_c\right) \bar P_0^{\cal C}\,,\nn
    0 &= \Gamma_w^{\cal C} (1+n_w) \bar P_2^{\cal C} + \Gamma_c^{\cal C} n_c \bar P_0^{\cal C}\nn
    &\qquad- \left(\Gamma_w^{\cal C} n_w + \Gamma_c^{\cal C} (1+n_c)\right) \bar P_1^{\cal C}\,,\nn
    0 &= \Gamma_w^{\cal C} n_w \bar P_1^{\cal C} + \Gamma_h^{\cal C} n_h \bar P_0^{\cal C}\nn
    &\qquad- \left(\Gamma_w^{\cal C} (1+n_w) + \Gamma_h^{\cal C} (1+n_h)\right) \bar P_2^{\cal C}\,,
\end{align}
where $\Gamma_\nu^{\cal C}$ are cycle-dependent as they are given by the bare $\Gamma_\nu$ rates multiplied with the squared Clebsch-Gordan coefficients from Eq.~\eqref{EQ:ladder_coefficients}.
Analogous equations can be written down for the negatively-oriented ($-$) cycles.
A solution to these equations exists, as these are just the steady-state conditions for a single QAR (see App.~\ref{APP:qar1}) without the normalization constraint. 

A fully analogous decomposition applies to the energy currents
\begin{align}
    \bar I_E^\nu &= \sum_{ij} (E_i-E_j) R_{ij}^\nu \bar P_j\nn
    &= \sum_{\cal C} \sum_{ij} (E_i-E_j) R_{ij}^{\nu,{\cal C}} \bar P_j^{\cal C}\,, 
\end{align}
and we get for the individual currents the cycle-resolved expressions (for positively oriented $+$ cycles)
\begin{align}\label{EQ:cyclecurrents}
    \bar I_E^c &= \delta \sum_{\cal C} \left[\Gamma_c^{\cal C} n_c \bar P_0^{\cal C} - \Gamma_c^{\cal C} (1+n_c) \bar P_1^{\cal C}\right]\,,\nn
    \bar I_E^h &= \Delta \sum_{\cal C} \left[\Gamma_h^{\cal C} n_h \bar P_0^{\cal C} - \Gamma_h^{\cal C} (1+n_h) \bar P_2^{\cal C}\right]\,,\nn
    \bar I_E^w &= (\Delta-\delta) \sum_{\cal C} \left[\Gamma_w^{\cal C} n_w \bar P_1^{\cal C} - \Gamma_w^{\cal C} (1+n_w) \bar P_2^{\cal C}\right]\,.
\end{align}
By eliminating one of the probabilities using~\eqref{EQ:steadystateplus} and comparing the terms in square brackets we can thus confirm the tight-coupling relations~\eqref{EQ:tightcoupling} in the main text.
Furthermore, the cooling condition $I_E^c > 0$ then also implies $I_E^w>0$ and $I_E^h<0$, such that from Eq.~\eqref{EQ:cyclecurrents} we get the conditions
$n_c \bar P_0^{\cal C} > (1+n_c) \bar P_1^{\cal C}$, $(1+n_h) \bar P_2^{\cal C} > n_h \bar P_0^{\cal C}$, and $n_w \bar P_1^{\cal C} > (1+n_w) \bar P_2^{\cal C}$.
Multiplying these conditions then eventually eliminates the dependence on the steady-state occupations
\begin{align}
    n_c(1+n_h) n_w > (1+n_c) n_h (1+n_w)\,,
\end{align}
which is equivalent to Eq.~\eqref{EQ:cooling_condition}.
Alternatively, Eq.~\eqref{EQ:cooling_condition} may also be obtained from the positivity of the entropy production rate $\bar\sigma_{\ii} = -\sum_\nu \beta_\nu \bar I_E^\nu \ge 0$ and the tight-coupling relations between the currents.
Departing from the tight-coupling limit will also alter the cooling condition~\cite{friedman2019a}.

\new{
\section{Stationary current in the non-collective limit}\label{APP:noncollective}

In the case that the stationary state is a classical one, i.e., a product state of mixtures without coherences
$\bar\rho = \bigotimes\limits_\ell \left[P_0^\ell \left(\ket{0}\bra{0}\right)_\ell + P_1^\ell \left(\ket{1}\bra{1}\right)_\ell+P_2^\ell \left(\ket{2}\bra{2}\right)_\ell\right]$, it is straightforward to see that the cooling current (compare Sec.~\ref{APP:fcs})
\begin{align}
    I_E^c &= \Gamma_c n_c \trace{S^c_+ S^c_- \bar\rho} - \Gamma_c (1+n_c) \trace{S^c_- S^c_+ \bar\rho}\nn
    &= \Gamma_c n_c \sum_{ij} h^c_i h^{c*}_j \trace{\left(\ket{1}\bra{0}\right)_i \left(\ket{0}\bra{1}\right)_j \bar\rho}\nn
    &\qquad-\Gamma_c (1+n_c) \sum_{ij} h^c_i h^{c*}_j \trace{\left(\ket{0}\bra{1}\right)_j \left(\ket{1}\bra{0}\right)_i \bar\rho}
\end{align}
will be additive in the number of qutrits: For such a steady state, in the above formula only the terms with $i=j$ can contribute under the trace, 
such that for $\abs{h^c_i}^2=1$ (compare red symbols in Fig.~\ref{FIG:currents}), the current would be linear in the number of qutrits $N$.
To the contrary, a superlinear scaling of the current indicates a deviation from this classical limit.

Numerically, we find that for 
$h^\nu_i = e^{\ii\varphi^\nu_i}$ as considered in the main text, the stationary state of Eq.~\eqref{EQ:lindblad} is close to the product state
\begin{align}\label{EQ:ssnoncoll}
    \bar\rho \approx \bigotimes\limits_\ell \left[P_0 \left(\ket{0}\bra{0}\right)_\ell + P_1\left(\ket{1}\bra{1}\right)_\ell+P_2 \left(\ket{2}\bra{2}\right)_\ell\right] 
\end{align}
with normalized probabilities $P_0+P_1+P_2=1$ corresponding to the steady-state solution of a single QAR.
Thus, we can link the near-linear scaling of the cooling current, which is what we observe in Fig.~\ref{FIG:currents} (red symbols and dashed red line), to the near product form of the stationary state, which would also be obtained for completely independent qutrits.  
}

\end{document}